\documentclass[10pt,twoside]{article}

\usepackage[T1]{fontenc}
\usepackage{newtxtext}
\usepackage{newtxmath}
\usepackage{graphicx}
\usepackage{amsmath}
\usepackage[authoryear]{natbib}
\bibpunct{(}{)}{;}{a}{}{,}
\usepackage{booktabs}
\usepackage{xcolor}
\usepackage{tikz}
\usetikzlibrary{shapes.geometric}
\usepackage{subcaption}
\usepackage[normalem]{ulem}
\usepackage{hyperref}
\hypersetup{colorlinks=true,urlcolor=blue,citecolor=black,linkcolor=black}
\usepackage{bookmark}

\makeatletter
\renewcommand\normalsize{%
  \@setfontsize\normalsize{10.7bp}{12pt}%
  \abovedisplayskip 6.5pt plus 1pt minus 1pt
  \belowdisplayskip \abovedisplayskip
  \abovedisplayshortskip 3pt plus 1pt
  \belowdisplayshortskip \abovedisplayshortskip}
\normalsize

\renewcommand\section{\@startsection{section}{1}{\z@}%
  {20pt plus 6pt minus 2pt}{3pt}{\fontsize{10.04}{12}\selectfont\bfseries}}
\renewcommand\subsection{\@startsection{subsection}{2}{\z@}%
  {12pt plus 3pt minus 3pt}{3pt}{\normalfont\normalsize\itshape\centering}}
\renewcommand\subsubsection{\@startsection{subsubsection}{3}{\z@}%
  {9pt plus 3pt minus 3pt}{3pt}{\raggedright\normalfont\normalsize\itshape}}
\renewcommand\paragraph{\@startsection{paragraph}{4}{\z@}%
  {9pt plus 3pt minus 3pt}{-3pt}{\normalfont\normalsize}}
\renewcommand\@seccntformat[1]{\normalfont\rmfamily\csname the#1\endcsname.\enskip}

\newcommand{\lefttitle}[1]{}
\newcommand{\righttitle}[1]{}
\newcommand{\aff}[1]{\textsuperscript{#1}}
\newcommand{\aff@inst}[1]{\par\vskip3pt\textsuperscript{#1}\unskip\ignorespaces}
\newcommand{\affiliation}[1]{\gdef\jfm@affiliation{\let\aff\aff@inst #1}}
\gdef\jfm@affiliation{}
\newcommand{\email}[1]{\href{mailto:#1}{#1}}
\newcommand{\corresau}[1]{\gdef\jfm@corresau{\textbf{Corresponding author:} #1}}
\gdef\jfm@corresau{}

\renewcommand\author[1]{\gdef\@author{\def\and{and }#1}}

\renewcommand{\maketitle}{%
  \newpage\global\@topnum\z@
  \vspace*{10pt}\addvspace{4.6pc}%
  \begingroup\raggedright
    {\fontseries{b}\fontsize{17bp}{19bp}\selectfont\@title\par}%
    \vspace*{23.5pt}%
    {\normalfont\fontsize{10.5bp}{14bp}\selectfont\bfseries\@author\par}%
    \vspace*{2pt}%
    {\normalfont\small\jfm@affiliation\par\vspace*{2pt}}%
    {\normalfont\small\jfm@corresau\par\vskip7pt}%
  \endgroup}

\renewenvironment{abstract}
  {\par\vspace*{33pt}\hrule\vspace*{11.75pt}\normalfont\normalsize\noindent\ignorespaces}
  {\par\vskip 9pt plus 1pt minus 1pt}

\newenvironment{appen}
  {\par\begingroup\appendix\footnotesize
   }
  {\endgroup}
\makeatother

\usepackage[format=plain,margin=0pt,width=\textwidth,justification=justified,
  labelfont=small,textfont=footnotesize,labelsep=period,singlelinecheck=true]{caption}

\lefttitle{A. Veilleux, H. Deniau and O. Vermeersch}
\righttitle{Journal of Fluid Mechanics}

\title{Bypass transition under wall cooling over a flat plate with an elliptical leading edge}
\author{A. Veilleux\aff{1}, H. Deniau\aff{1} \and O. Vermeersch\aff{1}}
\affiliation{\aff{1}ONERA/DMPE, Universit\'e de Toulouse, F-31055 Toulouse, France}
\corresau{A. Veilleux, \email{adele.veilleux@onera.fr}}

\begin{document}
\maketitle

\begin{abstract} 
This study uses wall-resolved large-eddy simulations to investigate bypass transition in a flat-plate boundary layer subjected to free-stream turbulence. Two thermal configurations are considered: an adiabatic wall and a uniformly cooled wall. The plate geometry includes an elliptic leading edge and is designed for direct experimental reproduction, providing a framework for future numerical-experimental comparisons.
The resolved leading edge allows direct examination of the early stages of bypass transition. Both simulations recover the classical sequence of receptivity, vortex tilting, lift-up, streak amplification, secondary instability and turbulent-spot growth. The wall-normal transport term is identified as a precursor of streak formation, while shear sheltering is quantified and linked to the frequency-dependent penetration of free-stream disturbances into the boundary layer.
Wall cooling does not modify the bypass-transition mechanisms nor the onset location of transition. Instead, it generates thermal streaks alongside the velocity streaks and shifts the latter slightly closer to the wall, consistent with optimal-perturbation predictions. The velocity streaks retain similar amplitudes, growth rates and spanwise spacings in both thermal conditions.
A conditional analysis reveals a pronounced asymmetry between high- and low-velocity streaks. Although breakdown is systematically observed within low-velocity streaks, the strongest pre-transitional evolution occurs within the high-velocity streak population, which undergoes significant amplification and a progressive displacement towards the wall before the onset of intermittency. These observations suggest that high-velocity streaks may actively contribute to the reorganisation of the streak field leading to secondary instability and breakdown.
\end{abstract}

\section{Introduction}
\label{sec:intro}
The transition from laminar to turbulent boundary layer flow is a critical phenomenon in many aerodynamic systems, particularly in turbomachinery. It significantly influences aerodynamic losses, heat transfer, and component lifespan due to thermal and mechanical stresses. In high-pressure turbines, elevated levels of free-stream turbulence (FST) originating from upstream stages favor bypass transition as the dominant mechanism \citep{Mayle1991, Wu1999}. Unlike natural transition dominated by the modal amplification of Tollmien–Schlichting waves, bypass transition is initiated by the receptivity of the boundary layer to free-stream disturbances. 

\citet{Morkovin1969} introduced the term bypass to describe any transition route that circumvents the classical TS-wave-driven process. In typical bypass transition scenarios, elongated streamwise velocity perturbations, sometimes referred to as Klebanoff modes \citep{Klebanoff1971, Wu2001}, emerge and amplify \citep{Westin1994}, eventually undergoing secondary instabilities that lead to the formation of turbulent spots \citep{Matsubara2001}. The term bypass transition has since become associated with this specific breakdown mechanism under high levels of FST \citep{Zaki2013}. 

\subsection{Bypass transition mechanisms}
The initial phase of bypass transition is the receptivity stage, during which vortical free-stream disturbances interact with the laminar boundary layer. The spectral content of the FST plays a critical role in determining which perturbations can penetrate the near-wall region. Due to the shear sheltering effect, high-frequency disturbances are strongly attenuated and their penetration depth rapidly decreases with increasing frequency, whereas low-frequency components can effectively reach and influence the boundary layer \citep{Jacobs1998,Hunt1999}. Experimental measurements by \citet{Hernon2007} further confirmed this behaviour by showing that the penetration depth scales inversely with disturbance frequency. By explicitly including the leading edge of the flat plate in the numerical domain, \citet{Nagarajan2007} showed that the vortical structures present in the free-stream upstream of the leading edge wrap around it and reorient into streamwise vortices. These vortices then undergo stretching as they enter the boundary layer, resulting in a localized amplification of the streamwise vorticity component $\omega_x$. \citet{Goldstein1998} used rapid distortion theory to explain how vortex stretching around the leading edge amplifies low-frequency free-stream disturbances and generates streamwise vorticity within the boundary layer. This mechanism is consistent with the later shear-sheltering interpretation of bypass transition, reinforcing the preferential penetration and growth of long-wavelength disturbances inside the boundary layer.

Once these low-frequency vortical disturbances enter the boundary layer, they induce vertical displacements of fluid: low-momentum fluid is lifted away from the wall, while high-momentum fluid is swept downward, producing alternating high- and low-velocity streaks. This lift-up mechanism \citep{Landahl1980} is responsible for the strong amplification of elongated streamwise streaks. The lift-up effect is linear and non-modal in nature, and the growth of streaks is well described within the framework of transient growth theory. Transient growth analyses \citep{Butler1992, Andersson1999, Luchini2000} identify the optimal perturbations, i.e. those leading to maximum energy amplification, and accurately predict the spatial structure of streaks observed in both experiments \citep{Matsubara2001,Nolan2012} and simulations \citep{Jacobs2001,Brandt2004, Nagarajan2007}.

Following their linear amplification, streaks undergo nonlinear deformation and may become susceptible to secondary instabilities. A nonlinear lift-up effect has been reported by \citet{Mao2017}, in which low-velocity streaks are further displaced toward the boundary-layer edge while high-velocity streaks are pushed toward the wall. Amplified streaks may then develop secondary inflectional instabilities, leading to local breakdown and the onset of turbulence. The link between secondary streak instabilities and breakdown has been supported by both numerical \citep{Brandt2004, Schlatter2008, Hack2014a} and experimental investigations \citep{Asai2002, Mans2007}.

Secondary instabilities often manifest as high-frequency oscillations localized on the streaks and can take the form of either sinuous or varicose modes, depending on the spanwise symmetry of the disturbance \citep{Swearingen1987}. Sinuous instabilities are antisymmetric and typically associated with spanwise oscillations of lifted low-velocity streaks, whereas varicose instabilities are symmetric and develop closer to the wall, linked to strong wall-normal shear layers surrounding the streaks. These instabilities can also be classified as inner or outer modes depending on their wall-normal location \citep{Vaughan2011}, with outer modes generally associated with lifted streaks near the edge of the boundary layer and inner modes with instabilities close to the wall. Numerical and experimental studies alike have confirmed the central role of these secondary instabilities in the onset of turbulence, as they frequently precede the emergence of turbulent spots, signaling the final stage of the transition process.

Recent comprehensive reviews of the destabilisation mechanisms can be found for example in \citet{FaundezAlarcon2024}, which highlights the ongoing debate regarding the precise role of streak secondary instabilities in bypass transition. Alternative mechanisms have been proposed, including localized near-wall wavepackets \citep{Nagarajan2007} or breakdown scenarios exhibiting similarities with natural transition \citep{Wu2017, Wu2023}, rather than classical sinuous or varicose streak instabilities \citep{Brandt2008}. The local stability analysis by \citet{FaundezAlarcon2024} confirms the localized nature of unstable modes, which are consistently observed near lifted low-velocity streaks. The authors also find no evidence of classical TS-wave instabilities for the studied FST spectrum. Their results therefore reinforce the view that, under bypass-transition conditions induced by FST, streak breakdown through secondary instability mechanisms, rather than classical TS-wave amplification, constitutes the primary route to transition.

The breakdown of streaks ultimately leads to the formation of turbulent spots, i.e. localized regions of turbulence that grow, merge, and progressively fill the boundary layer. 
Spot formation has been observed experimentally \citep{Matsubara2001, Mans2005} and numerically \citep{Jacobs2001, Brandt2004}. The leading edge of a spot travels near the free-stream velocity, while the trailing edge moves more slowly, causing elongation and eventual merging.

\subsection{Explored Parameters in Bypass Transition Studies}
The bypass transition to turbulence has been the subject of extensive investigation through both experimental and numerical approaches, with numerous studies exploring the influence of key parameters governing the transition process.

Among these, the FST has received the most attention. It plays a pivotal role in modulating the dynamics of streak formation and breakdown. Elevated FST intensities are known to accelerate the amplification of streaks and hasten the onset of transition, with pronounced effects on streak amplitude, spanwise spacing, and transition location. Beyond turbulence intensity, the influence of other FST characteristics, including the integral length scale, isotropy, and spectrum, has been extensively investigated through experiments \citep{Klebanoff1971,Westin1994,Jonas2000,Matsubara2001,Fransson2005} and numerical simulations \citep{Brandt2004,Ovchinnikov2008,Zaki2010}. In particular, larger integral length scales have been associated with earlier transition and longer transitional regions, indicating that turbulence intensity alone is insufficient to characterize the transition process. More recently, \citet{Fransson2020} showed that the effect of the integral length scale can reverse depending on the FST conditions, and proposed a scale-matching mechanism based on an optimal ratio between the FST integral scale and the boundary-layer thickness at transition. DNS of the same experimental configuration by \citet{Durovic2024} supported this non-monotonic dependence on the integral length scale.

The role of the leading edge in bypass transition has also been emphasized in several numerical studies. \citet{Ovchinnikov2008} investigated the effect of the domain configuration by comparing simulations with a full-domain mesh and a symmetry-split mesh, both using a super-elliptic leading edge (aspect ratio $AR = 6$). Their results show that the symmetry condition at the leading edge inhibits vertical velocity fluctuations ($v'=0$), preventing the generation of streamwise disturbances $u'$ near the stagnation point. As a consequence, the amplitude of streamwise fluctuations is reduced, and the onset of transition is delayed in the split-domain configuration, highlighting the importance of resolving the full leading-edge geometry to accurately capture receptivity. Using a wall-resolved LES approach, \citet{Nagarajan2007} further demonstrated that increasing the leading-edge bluntness (from $AR = 6$ to $10$) causes both the onset and completion of transition to move upstream, underscoring the sensitivity of the transition process to the leading-edge shape. This finding is consistent with the results of \citet{Wang2019}, who performed linear and nonlinear stability analyses and showed that the presence of the leading edge enhances the generation of streamwise vorticity. This, in turn, promotes streak amplification via the lift-up effect and reinforces the importance of leading-edge receptivity in the early stages of bypass transition.

Finally, the effect of streamwise pressure gradients has also been extensively addressed. An adverse pressure gradient tends to enhance streak amplitudes and promote earlier transition, while a favourable pressure gradient suppresses disturbance growth and thus delays transition \citep{AbuGhannam1980, Zaki2006, Nolan2013}.More recently, \citet{Mamidala2022} highlighted the role of leading-edge pressure gradients in boundary-layer receptivity to FST, while \citet{Zhao2020} investigated the combined influence of pressure gradients, surface curvature, and FST intensity on bypass transition in a turbine-relevant configuration.

Despite its important role in engineering applications, the influence of wall thermal conditions in bypass transition remains relatively underexplored. From a theoretical perspective, \citet{Tumin2003} analysed the transient growth of three-dimensional stationary perturbations in a compressible boundary layer using a non-parallel linearised boundary-layer formulation. Their results show that the optimal disturbances remain steady streamwise vortices generating streaks through the lift-up mechanism for different wall temperatures, while the wall temperature strongly affects the thermal and velocity structure of the perturbations, including the formation of temperature streaks. In particular, wall cooling significantly enhances the transient energy growth. This behaviour was further clarified by \citet{Vermeersch2009Thesis}, who used optimal perturbation theory (OPT) to show that, in subsonic flows, wall cooling does not significantly modify streamwise velocity fluctuations, but instead generates stronger temperature fluctuations within the boundary layer, leading to a substantial increase in the overall perturbation energy gain. In contrast, supersonic cases exhibit a different behaviour, where reduced temperature fluctuations are compensated by stronger velocity perturbations, resulting in comparatively small variations of the total growth.

Experimentally, early investigations by \citet{Sohn1991} on a heated flat plate suggested that wall heating may influence bypass transition through modifications of the thermal boundary layer and enhanced low-frequency unsteadiness. However, the opposite scenario of wall cooling, more relevant to high-pressure turbine blades, was not addressed. Wall cooling was experimentally investigated by \citet{Rued1986}, who examined a boundary layer subjected to both FST and favourable pressure gradients. Although the term \emph{bypass transition} was not explicitly used, the investigated conditions clearly correspond to transition induced by elevated FST. The authors observed that wall cooling (down to a temperature ratio $T_{\infty}/T_{wall} = 1.89$) had only a limited influence on the transition location compared with the adiabatic case at elevated turbulence levels. Instead, wall cooling primarily enhanced the near-wall thermal gradients, leading to increased heat transfer in both laminar and turbulent regimes. Their results therefore suggest a strong thermal modulation of the mean boundary-layer properties, rather than major modifications of the transition process itself.
More recently, \citet{Ferreira2019} conducted an experimental investigation of bypass transition in a high-pressure turbine cascade, exploring the impact of varying gas-to-wall temperature ratios ($T_{\infty}/T_{wall}$ from 1.14 to 1.51) under different turbulence and Mach number conditions. For the low-Mach-number cases, especially at elevated turbulence levels ($Tu = 5.3\%$), they reported a modest increase in transition length with increasing temperature ratios, while no clear shift in transition onset was observed.
Using the same turbine configuration, \citet{Rubini2018} investigated the ability of RANS transition models to reproduce thermal effects in bypass transition. Their study highlighted the difficulty of current modelling approaches in accurately capturing the influence of thermal boundary conditions on the transition process, even when locally based formulations were employed. These findings underline the current limitations of RANS transition models in the presence of aero-thermal coupling and reinforce the need for high-fidelity simulations and experimental databases to support future model development and validation.

\subsection{Motivation of the Present Work}

Bypass transition has been widely studied under a variety of aerodynamic and disturbance conditions. However, the specific influence of wall cooling on its development remains an open question, particularly in configurations representative of high-pressure turbine blades, where the free-stream temperature is significantly higher than that of the wall and the incoming turbulence intensity is elevated. In such conditions, it is still unclear how thermal conditions affect the transition mechanisms—from receptivity to breakdown—when the boundary layer is subjected to elevated levels of FST.
To address this gap, we consider an academic yet realistic configuration and conduct two LES over a flat plate with an elliptical leading edge: one with an adiabatic wall, and one with a uniform wall cooling at a temperature ratio \(T_{\infty}/T_{wall}=2.0\), representative of high-pressure turbine vane environments. 

To reduce computational cost, several numerical studies have deliberately excluded the leading edge from the computational domain, starting sufficiently far downstream so that a Blasius boundary layer is already established, and then introducing synthetic FST at the inlet. In this framework, FST can be generated via the continuous spectrum of the Orr-Sommerfeld and Squire operators in the wall-normal direction, as in the DNS of \citet{Jacobs2001} and \citet{Brandt2004}. To better reproduce the interaction between FST and the boundary layer, \citet{Pinto2019} proposed injecting synthetic turbulence above the boundary layer rather than directly within it. Their approach reproduced the ERCOFTAC T3A test case but highlighted that the transition behaviour remains sensitive to the manually prescribed structure and wall-normal location of the injected disturbances. However, the absence of a leading edge prevents the receptivity process from being fully resolved and does not explicitly account for the mechanisms governing the initial formation of streamwise streaks.

In contrast, the present study includes the full plate geometry, allowing disturbances to develop naturally from the stagnation region under the action of incoming FST. The configuration was designed to remain representative of future experiments while preserving the key physical mechanisms governing receptivity and bypass transition. FST is imposed upstream of the leading edge so that receptivity, streak formation and subsequent bypass transition arise naturally within the computational domain. Particular attention is devoted to assessing how wall cooling affects the streak dynamics and the transition process, from the earliest stages of receptivity to the onset of breakdown.

The paper is organised as follows. Section~2 details the numerical methodology, including the computational setup, boundary conditions, turbulence generation and validation metrics. Results are presented in Section~3. After characterising the incoming FST and the global boundary-layer response, the receptivity process, streak formation and breakdown mechanisms are analysed in detail. The influence of wall cooling on velocity and thermal streaks is discussed and compared with OPT predictions. Particular attention is given to the evolution of high- and low-velocity streaks and to their respective roles in the transition process. Finally, discussion and concluding remarks are provided.

\section{Simulation details}
\subsection{Geometry, flow conditions and numerical setup}
The configuration considered is that of a flat plate with an elliptical leading edge of aspect ratio $AR = 10$, defined by the semi-minor and semi-major axes $r=15$~mm and $R=150$~mm, respectively. The plate thickness is set to $30$~mm, which ensures both mechanical integrity and sufficient space for the integration of a cooling device in future experiments. The overall plate length is $L = 208.9$~mm, corresponding to a normalized streamwise extent of only $L/R \approx 1.39$ when compared with the DNS configuration of \citet{Nagarajan2007}, whose computational domain extended up to approximately $x/R \simeq 7$. The present setup is therefore expected to primarily capture the receptivity and early bypass-transition stages, with only a limited region of fully developed turbulence near the plate end. The computational domain spans a transverse width of $L_z = 15.8766$~mm, chosen to ensure an adequate spanwise resolution while capturing approximately five to ten streaks across the span at a reasonable computational cost.
The free-stream Mach number is $M=0.1$ and the static temperature is $T_\infty = 293.15$~K. These conditions yield a leading-edge Reynolds number of $Re_r = 34\,000$, i.e.\ at least one order of magnitude higher than the values typically employed in previous numerical investigations of bypass transition (e.g.\ \citep{Jacobs2001, Ovchinnikov2008}). Two LES configurations are considered: an adiabatic wall (hereafter denoted AW) and a cooled wall (CW), the latter corresponding to a wall temperature ratio $T_\infty/T_{wall} = 2.0$. 

The simulations are performed with the ONERA \textit{elsA} flow solver \citep{Cambier2002}, which integrates the compressible Navier--Stokes equations on structured multi-block grids using a finite-volume formulation. Spatial discretization relies on a low-dissipation, low-dispersion compact scheme originally proposed by \citet{Lele1992} and adapted to finite volumes by \citet{Fosso2011}. This high-order scheme (up to $5^{\mathrm{th}}$ order) enables accurate resolution of wavelengths sampled with as few as 7--10 grid points. To ensure numerical stability and provide appropriate subgrid-scale dissipation, an explicit $8^\text{th}$-order filtering procedure following \citet{Visbal2002} is applied. The overall methodology therefore amounts to an implicit LES. Time integration is carried out using a six-stage explicit Runge--Kutta algorithm \citep{Bogey2004}, stable for Courant numbers up to 1.1. To initialize the LES and define consistent boundary conditions, a two-dimensional steady RANS simulation is first performed over an extended domain. The resulting laminar base flow provides the mean velocity and thermodynamic fields required for the inflow and farfield boundaries of the LES. A subdomain is extracted from the RANS solution and extruded in the spanwise direction to generate the three-dimensional computational domain for the LES. This volume is subsequently re-meshed with wall-resolved resolution to satisfy LES specific grid constraints. The initial condition for the three-dimensional simulation corresponds to the extruded RANS solution, ensuring a physically consistent and divergence-free starting point. Due to the explicit time integration and the fine near-wall grid spacing, the physical time step is constrained to $\Delta t = 3.89 \times 10^{-8}$~s. After a transient initialization phase, required to purge the initial conditions and allow turbulent structures to populate the computational domain, the main LES was run for $5.38 \times 10^{-2}$~s of simulated time for each wall condition.

\subsection{Computational domain, grid and boundary conditions}
The computational mesh contains $2401\times 501\times 325$ points in $(x,y,z)$, i.e.\ $\approx 3.89\times 10^{8}$ cells. Periodic boundary conditions are applied in the spanwise direction. To faithfully capture both the proper behaviour of the incoming turbulence and the transition onset, the computational domain extends sufficiently far upstream and downstream of the plate. A two-dimensional view of the grid (displayed with one point out of 16), together with the boundary conditions and probe distribution, is illustrated in figure~\ref{fig:LESmesh}. In this and subsequent figures, the streamwise and wall-normal coordinates are normalised as $x/L$ and $y/r$. 
The outlet (blue) employs Navier--Stokes characteristic boundary conditions (NSCBC) with pressure relaxation \citep{Poinsot1992}. The wall (black) is either adiabatic or isothermal depending on the configuration. Along the top boundary, a non-reflecting condition following \citet{Tam1996} is imposed, with the reference state taken from the base flow of the preliminary RANS simulation (orange). The inflow condition (green) is constructed by superimposing turbulent fluctuations onto this preliminary RANS profile, enabling a natural receptivity and transition process. External turbulence is introduced at this boundary using the random Fourier mode method of \citet{Bechara1994}, based on the von Kármán energy spectrum \citep{vonKarman1948}. A total of 100 modes are used, spanning integral length scales from $l_{\min} = 1.5 \times 10^{-4}$~m to $l_{\max} = 10^{-2}$~m. Each inflow forcing period spans 7520 iterations ($\approx 2.9\times10^{-4}$ s), corresponding to the advection time of the largest turbulent structure through the domain, after which a new realization is generated. 
This approach provides a continuously regenerated turbulent inflow and was calibrated to obtain statistically stationary turbulence levels at the inlet.
Prior to the flat-plate simulations, an auxiliary LES was carried out in a
homogeneous rectangular box using the same synthetic-turbulence forcing, target
energy spectrum and numerical solver as in the present study. This reference
configuration, used to assess the isotropy of the forcing independently of the
flat-plate geometry, is described in Appendix~\ref{app:cube}.

\subsection{Probes and data outputs}
\begin{figure}
  \centerline{\includegraphics[scale=1]{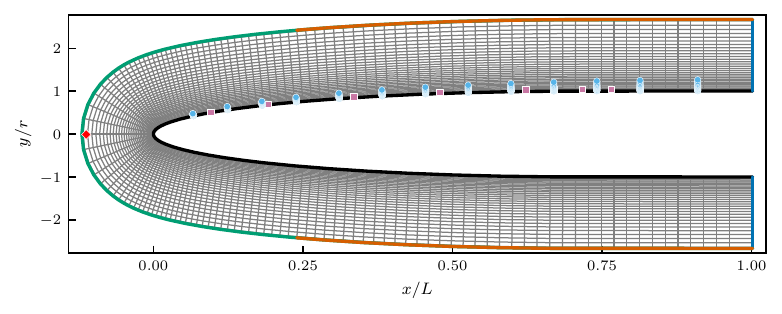}}
  \caption{Coarsened 2D view of the structured grid ($1/16$ point) around the flat plate with an elliptic leading edge. Boundary conditions: wall (black); outflow NSCBC (blue); inflow with synthetic-turbulence injection superimposed to the RANS solution (green); boundary as RANS solution (orange). Monitoring probes are also displayed for the AW case: spanwise lines positions (pink squares); wall-normal profiles (blue circles); inflow probe (red diamond).}
\label{fig:LESmesh}
\end{figure}

Three types of flow data are recorded during the LES. 
First, time-resolved signals are collected through a system of probes deployed within the domain, as shown in figure~\ref{fig:LESmesh}. Nine lines of 80 probes, spanning the $z$ direction at mid-boundary-layer height (pink squares), allow assessment of the spanwise coherence of streaks. Thirteen wall-normal profiles, each composed of 19 probes (blue circles) at $z=-1$~mm, are positioned at several streamwise locations to capture boundary-layer development and the shear-sheltering mechanism. Injected turbulence is monitored via three probes ($z = {-3.75, 0, 3.75}$~mm) located upstream of the plate near the inflow boundary (red diamond). Probe signals are sampled at a frequency of approximately $2.6~\text{MHz}$. Then, first- and second-order turbulence statistics, including mean velocity components and Reynolds-stress components, are accumulated on the fly throughout the simulation. Finally, instantaneous three-dimensional flow fields are stored at a sampling frequency of $3.42~\text{kHz}$.
To account for the elliptic leading edge, variables close to the plate are expressed in a local curvilinear basis attached to the plate geometry, with $\parallel$ and $\perp$ denoting the tangential and normal directions to the wall, respectively, and $z$ the spanwise direction. Details on data processing and on the definition of the plate-attached frame are provided in Appendix~\ref{app:post}.

\subsection{Grid resolution in wall units}
\begin{figure}
  \centerline{\includegraphics[scale=1]{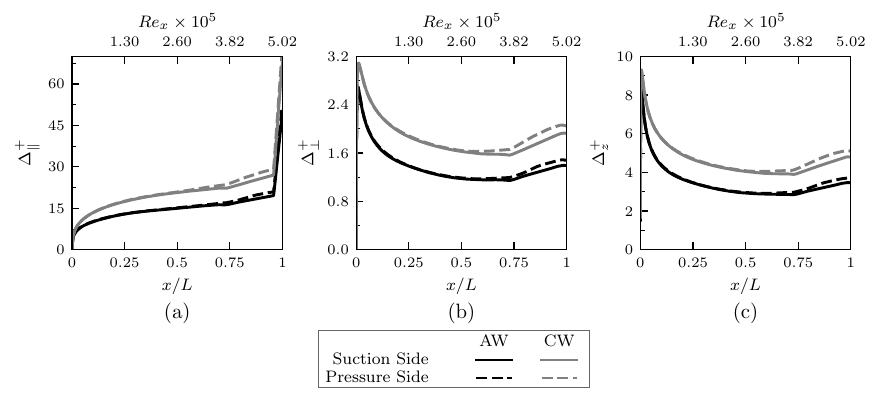}}
  \caption{Streamwise evolution of the grid resolution in wall units: (\textit{a}) tangential spacing $\Delta_{\parallel}^+$; (\textit{b}) wall-normal spacing $\Delta_{\perp}^+$ (first off-wall spacing); and (\textit{c}) spanwise spacing $\Delta_z^+$. 
  }
\label{fig:LES_dplus}
\end{figure}

Beyond the overall mesh layout and boundary conditions, it is essential to ensure that the grid resolution meets accepted wall-resolved LES requirements. To this end, figure~\ref{fig:LES_dplus} reports the evolution of the wall-unit spacings along the plate.  
In the following, spacings are expressed in the curvilinear frame attached to the wall, as $\Delta_{\parallel}^+$ (tangential), $\Delta_{\perp}^+$ (normal), and $\Delta_z^+$ (spanwise). For clarity, the abscissa is shown both as the streamwise coordinate normalised by the plate length ($x/L$, bottom axis) and as the corresponding Reynolds number $Re_x$ (top axis).
For both AW and CW cases, the pressure- and suction-side profiles closely overlap over most of the plate and exhibit only small differences in the downstream region. As shown later, these differences coincide with the onset of transition and the associated rise in local skin friction.
As expected, the CW case exhibits larger $\Delta^+$ values than the AW case, with a typical CW/AW ratio of $\approx 1.3$, reflecting the higher friction velocity in the cooled boundary layer. For the tangential spacing, $\Delta_{\parallel}^+<20$ in AW (and $<30$ in CW). The first off-wall spacing, $\Delta_{\perp}^+$, peaks immediately downstream of the nose at $\max \Delta_{\perp}^+\approx 2.7$ for AW (resp.\ $\approx 3.1$ for CW) and then remains within $1$--$2$ for AW (resp.\ $1.5$--$2.5$ in CW). The spanwise spacing reaches $\max \Delta_z^+\approx 8.5$ just downstream of the leading edge in AW (resp.\ $\approx 9$ in CW), and subsequently decreases to $3$--$5$ over most of the plate in AW (resp.\ $4$--$6$ in CW).  

These values are consistent with established LES guidelines. In particular, the use of a high-order compact discretization allows a relaxed streamwise constraint of $\Delta_{\parallel}^+\approx 40$ \citep{Nagarajan2007}, whereas second-order formulations typically require $\Delta_{\parallel}^+\approx 10$ \citep{Jacobs2001, Ovchinnikov2008, Brandt2004}; in our computations $\Delta_{\parallel}^+$ remains safely below $30$ nearly up to the plate end. In the wall-normal direction, the conventional $y_1^+<1$ condition is slightly relaxed to maintain computational feasibility, while in the spanwise direction $\Delta_z^+$ lies within the $3.5$--$10$ range commonly employed in transitional and turbulent boundary-layer LES \citep{Muthu2020}. Taken together, these metrics support the adequacy of the grid for wall-resolved LES of bypass transition in both AW and CW configurations.

\section{Results and analysis}
The two LES are now analysed, with emphasis on the effect of wall thermal conditions on the boundary-layer transition process. This section is organised in two parts: first, the FST  and its imprint on the boundary layer are characterised; then, the mechanisms underlying bypass transition over the plate are analysed. 

  \subsection{Injected perturbations and boundary layer response}
\begin{figure}
          \centerline{\includegraphics[scale=1]{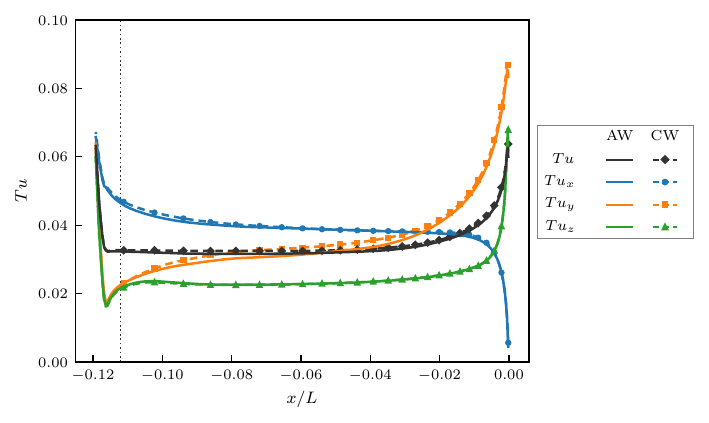}}
            \caption{Streamwise evolution of the turbulence intensity $Tu$  and its components along the mid-plane ($y=0$) upstream of the flat plate. The vertical dotted line marks the free-stream probe location.}
            \label{fig:Tu_upstream}
        \end{figure}

Before examining the onset and development of transition, the imposed FST and its evolution from the inflow boundary to the leading-edge region are first characterised.
Figure~\ref{fig:Tu_upstream} reports the streamwise evolution of the turbulence
intensity upstream of the plate, evaluated along the mid-plane at $y=0$. The turbulence intensities are computed from the turbulence statistics and averaged in the spanwise direction. The total turbulence intensity,
\[
Tu=\frac{\sqrt{(u_{\mathrm{rms}}^2+v_{\mathrm{rms}}^2+w_{\mathrm{rms}}^2)/3}}{U_\infty},
\]
(black, diamonds) is decomposed into its streamwise, vertical and spanwise contributions,
\[
Tu_x=\frac{u_{\mathrm{rms}}}{U_\infty}, \qquad
Tu_y=\frac{v_{\mathrm{rms}}}{U_\infty}, \qquad
Tu_z=\frac{w_{\mathrm{rms}}}{U_\infty},
\]
shown respectively in blue (circles), orange (squares) and green (triangles). Results are displayed for both the adiabatic-wall (AW, solid lines) and cooled-wall (CW, dashed lines) cases. The vertical dotted line indicates the probe position used as reference for the inflow turbulence level.

The two simulations, AW and CW, exhibit identical turbulence levels upstream of the plate, as expected since the wall temperature does not influence the inflow forcing and no temperature fluctuations were imposed. While thermal fluctuations may be relevant in turbine environments, they are beyond the scope of the present study. The following discussion therefore applies to both cases. At the inflow boundary, the total turbulence intensity is injected at a level slightly above $6\%$, with equal contributions from the streamwise, vertical and spanwise components, indicating an initially balanced distribution of turbulent kinetic energy among the three velocity components. Over a short downstream distance, the turbulence intensity decreases sharply and settles around $3.5\%$. This decay is primarily associated with the vertical and spanwise components, which rapidly fall to approximately $2\%$, whereas the streamwise component remains close to $5\%$. This adjustment produces an anisotropic FST field dominated by streamwise fluctuations. Consequently, although the inflow disturbances are initially imposed at a level slightly above $6\%$, the boundary layer effectively develops under a FST level of $Tu \simeq 3.5\%$. This value lies within the range commonly adopted in studies of bypass transition and is comparable to the FST levels considered in previous numerical \citep{Brandt2004,Nagarajan2007} and experimental investigations \citep{Jonas2000,Nolan2012}.
Further downstream, in the region unaffected by the plate ($x/L \lesssim -0.025$), the total turbulence level remains close to $3.5\%$. The spanwise contribution stays nearly constant, while the vertical component slowly increases and the streamwise one decreases, until both converge to $\approx 4\%$. This crossover marks the end of the purely free-stream evolution. As the flow approaches the leading edge, the presence of the plate alters the turbulence distribution: the total intensity increases, driven by a rise in both the spanwise and vertical components. The latter grows more rapidly, reflecting the conversion of streamwise fluctuations into vertical fluctuations induced by the deflection of incoming turbulent structures over the elliptic nose of the plate, as further illustrated in figure~\ref{fig:vortex_tilting} and discussed in section~\ref{sec:receptivity}.

\begin{figure}
          \centering 
        \includegraphics[scale=1]{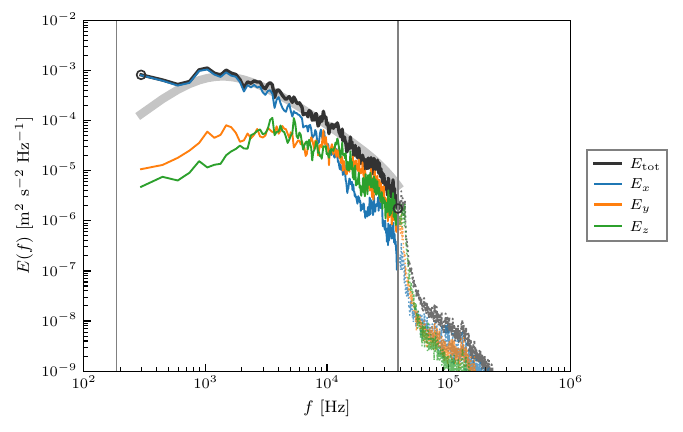}
        \caption{
        Energy spectra of the velocity fluctuations upstream of the flat plate leading edge ($x=-2.3$~cm) for the CW simulation. 
        Contributions from the three velocity components and their sum $E_{\mathrm{tot}}$ are shown. 
        Solid lines denote the resolved frequency range, bounded by $f_{\min}$ and $f_{\max}$ (vertical solid lines). 
        Dotted extensions indicate unresolved frequencies, and open circles mark the resolved-band endpoints. 
        The grey line denotes the theoretical spectrum obtained from Pope's model.}
      \label{fig:spectrum_Pope_PF}
      \end{figure}

Figure~\ref{fig:spectrum_Pope_PF} presents the energy spectra of the
injected turbulence for the CW simulation, evaluated from the velocity signals
recorded at the three free-stream probes upstream of the plate (red diamond in Figure~\ref{fig:LESmesh}). At each probe,
the spectra of the $(x,y,z)$ components were computed and combined to form the
total energy spectrum $E_{\mathrm{tot}}$. The three probes were then averaged to
improve statistical convergence. The resulting spectra are compared with a
theoretical spectrum based on the model of \citet{Pope2000}, using the prescribed integral
length scale ($0.01$~m) together with the turbulent kinetic energy and mean
convection velocity measured at the probes. 
The resolved frequency band is bounded by $f_{\min}$ and
$f_{\max}$, with unresolved ranges shown as dotted extensions. The minimum
frequency is $f_{\min}=10/t_f=186$\,Hz, where $t_f$ is the total simulation
time, so that a wave is considered resolved if at least ten periods are
captured. The maximum frequency is set locally by a spatial-resolution
criterion: at each point, it corresponds to a wavelength resolved by eight grid
cells, using the local mean convection speed and the smallest grid spacing in
the mesh.

The total spectrum exhibits good agreement with the theoretical Pope spectrum
over most of the resolved frequency range. At the lowest resolved frequencies 
($f\lesssim 5 \cdot 10^2$~Hz), the total spectrum
slightly exceeds the theoretical Pope distribution. Given the limited number of
resolved periods in this frequency range, this departure may partly reflect
statistical uncertainty, together with the finite number of Fourier modes used
in the inflow forcing. In the
inertial range, the spectra follow closely the theoretical $-5/3$ slope up to
$f\approx3\times10^4$~Hz, beyond which the energy decays rapidly, reflecting
the effect of subgrid-scale dissipation.
The component-wise distribution confirms that the streamwise contribution
dominates over a broad frequency range ($f\lesssim10^4$~Hz), while the
wall-normal and spanwise components remain weaker. This anisotropy is
consistent with the turbulence-intensity evolution reported in
figure~\ref{fig:Tu_upstream}. It does not originate from the
synthetic-turbulence generation procedure itself, which recovers an isotropic
spectrum in a homogeneous reference configuration (Appendix~\ref{app:cube}), but rather from the present set-up. While the streamwise velocity fluctuations recover
closely the prescribed theoretical spectrum, the coarser wall-normal grid
spacing upstream of the plate ($\Delta y\approx2\Delta x$), together with the
limited spanwise width of the computational domain, restrict the
representation of the largest turbulent scales carried by the wall-normal and
spanwise velocity fluctuations. As a result, the prescribed free-stream
turbulence exhibits a preferentially streamwise character, although its overall
spectral distribution remains consistent with the target Pope spectrum. Nearly
identical spectra are obtained for the AW case and are therefore omitted for
brevity.

                \begin{figure}
          \centerline{\includegraphics[scale=1]{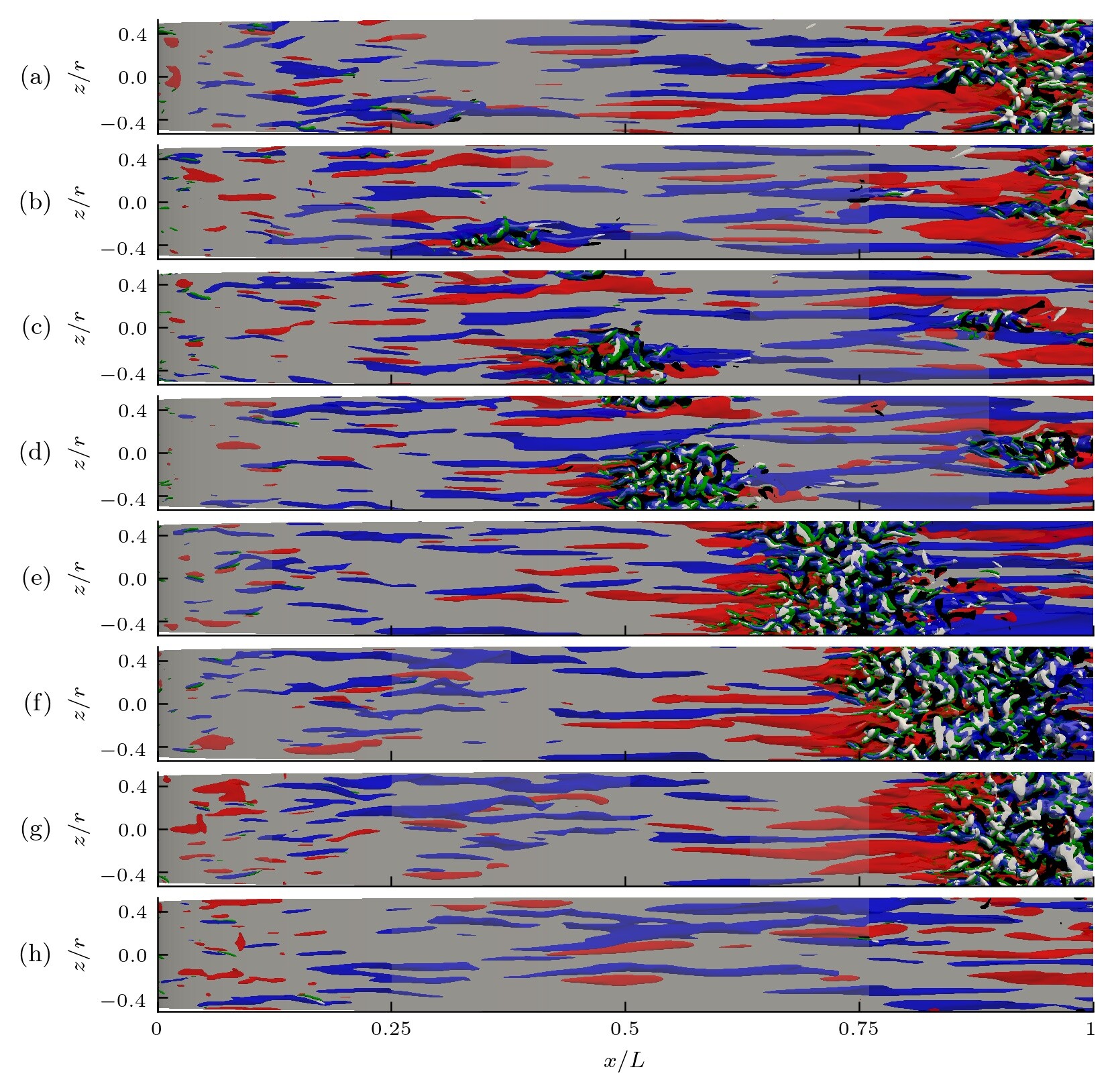}}
          \caption{Instantaneous flow fields in the $(x,z)$-plane at the lower surface of the plate for the AW case. Contours of $u'_{\parallel}=\pm 0.12\,U_\infty$ are shown in blue (negative) and red (positive), contours of $v'_{\perp}=\pm 0.05\,U_\infty$ in white (negative) and black (positive), and the non-dimensional $Q$-criterion $Q^\ast=Q\;r^2/U_\infty^2 =10$ in green. The snapshots correspond to $(t-t_0)/\Delta t_{\mathrm{snap}}=0,\,3,\,6,\,8,\,13,\,17,\,20,\,27$ for panels
(\textit{a})--(\textit{h}), respectively, where
$\Delta t_{\mathrm{snap}}=7520\,\Delta t_{\mathrm{LES}}
\simeq 2.9\times10^{-4}$~s.}

          \label{fig:flow_description}
        \end{figure}

        Having characterised the inflow turbulence and its upstream development, we now turn to the response of the boundary layer over the plate. Figure~\ref{fig:flow_description} shows instantaneous $(x,z)$-plane views of the lower surface for the AW case. Contours of the streamwise velocity fluctuation in the plate-attached frame $u'_{\parallel}=\pm 0.12\,U_\infty$ ($U_\infty = 36$~m/s being the free-stream velocity) are plotted together with contours of the normal velocity fluctuations $v'_{\perp}=\pm 0.05\,U_\infty$ and isosurfaces of the non-dimensional $Q$-criterion, $Q^\ast =Q\;r^2/U_\infty^2=10$. The formation of elongated streaks typical of bypass transition is immediately apparent. These streaks originate very close to the leading edge, and their streamwise extent increases progressively downstream.  In panel~(a), a low-velocity streak at $x/L=0.25$ exhibits a sinuous deformation, flanked by high-velocity streaks. It is separated from its neighbours by regions of wall-normal velocity fluctuations, and accompanied by vortical structures highlighted by the $Q^\ast$-criterion. The destabilisation of this streak is evident in panel~(b), and gives rise to a turbulent spot visible at $x/L=0.5$ in panel~(c).  Although only one representative breakdown event is shown here, the underlying streak destabilisation mechanisms and the statistical properties of the resulting turbulent spots are analysed separately in \S~\ref{sec:streak_desta_and_turbulent_spot}. The spot is characterised by the emergence of small-scale turbulent motions. As it convects downstream, it grows and spreads laterally (e), eventually contaminating the full span of the plate (f). The turbulent zone initiated by the spot is subsequently convected out of the computational domain (g), leaving behind a laminar boundary layer (h). This sequence highlights the local and intermittent nature of streak breakdown, consistent with the numerical observations of \citet{Jacobs2001}: not all streaks undergo destabilisation, and laminar regions can re-establish once a turbulent spot has been advected out of the domain.

\begin{figure}
          \centering
          \begin{subfigure}{0.49\textwidth}
            \includegraphics[scale=1]{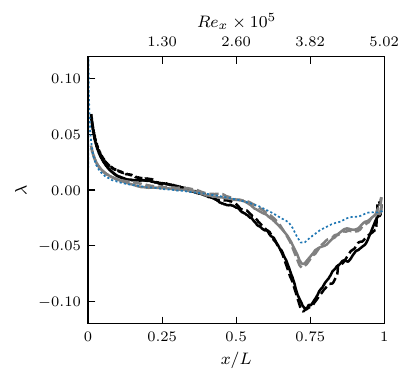}
            \caption{}
            \label{fig:lambda}
          \end{subfigure}\hfill
          \begin{subfigure}{0.49\textwidth}
            \includegraphics[scale=1]{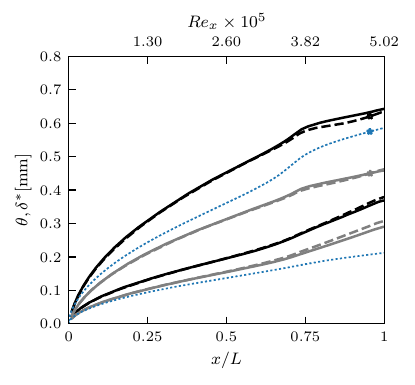}
            \caption{}
            \label{fig:theta_delta_star}
          \end{subfigure}\hfill
          \begin{subfigure}{0.49\textwidth}
            \includegraphics[scale=1]{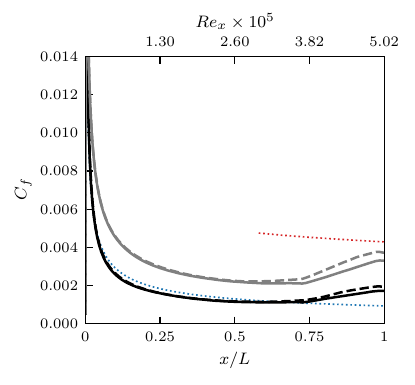}
            \caption{}
            \label{fig:Cf}
          \end{subfigure}\hfill
          \begin{subfigure}{0.49\textwidth}
            \includegraphics[scale=1]{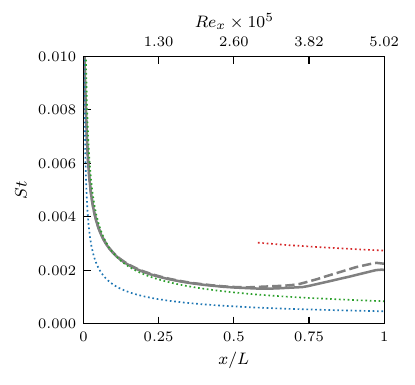}
            \caption{}
            \label{fig:stanton}
          \end{subfigure}

\vspace{0.5em}

\begin{center}
{\footnotesize
\setlength{\fboxsep}{3pt}
\setlength{\fboxrule}{0.3pt}
\fcolorbox{black!50}{white}{%
\begin{tabular}{c@{\hspace{0.8em}}c@{\hspace{1.0em}}c}
    & AW & CW \\[-0.2em]

Suction Side  &
\raisebox{0.3ex}{%
\tikz{\draw[line width=1.2pt] (0,0) -- (0.65,0);}
}
&
\raisebox{0.3ex}{%
\tikz{\draw[gray,line width=1.2pt] (0,0) -- (0.65,0);}
}
\\[-0.1em]

Pressure Side &
\raisebox{0.3ex}{%
\tikz{\draw[line width=1.2pt,dash pattern=on 6pt off 2pt] (0,0) -- (0.65,0);}
}
&
\raisebox{0.3ex}{%
\tikz{\draw[gray,line width=1.2pt,dash pattern=on 6pt off 2pt](0,0) -- (0.65,0);}
}

\end{tabular}%
}}
\end{center}

\vspace{0.2em}

          \caption{
          Global boundary-layer indicators along the flat plate for the AW and CW cases. 
          (\textit{a}) Pressure gradient parameter $\lambda = (\theta^2 / \nu) (\mathrm{d}U/\mathrm{d}x)$ with RANS laminar solution (blue).
          (\textit{b})  Momentum and displacement thicknesses $\theta$ and $\delta^*$ (star markers) with RANS laminar solution (blue). 
          (\textit{c}) Skin-friction coefficient $C_f$, with Blasius laminar solution (blue) and turbulent flat-plate correlation of Coles \& Fernholtz 1996 (red). 
          (\textit{d}) Stanton number $St$, with the laminar isothermal correlation (blue), its cold-wall correction $(T_{\infty}/T_{wall})^{0.85}$ (green; Kays \& Crawford 1993), and the turbulent correlation (red). }
          \label{fig:global_indic}
        \end{figure}

While the snapshots provide a qualitative picture of the intermittent nature of
bypass transition, it is important to examine how this behaviour
manifests in terms of global boundary-layer indicators. Figure~\ref{fig:global_indic} compares global boundary-layer indicators derived from the time-averaged turbulence statistics accumulated during the LES. 

Panel~(a) shows the Pohlhausen pressure-gradient parameter, $\lambda=(\theta^2/\nu)\,\mathrm{d}U/\mathrm{d}x$ for the AW and CW LES together with the same configuration simulated with a laminar RANS approach on an adiabatic wall. Here, $\theta$ denotes the momentum thickness. Near the leading edge all curves exhibit the same behaviour: a positive $\lambda$ associated with the favourable pressure gradient induced by flow acceleration around the elliptical nose, followed by a relaxation towards zero as the acceleration diminishes. Beyond $x/L \approx 0.4$, the three cases diverge. In the laminar RANS solution, $\lambda$ decreases steadily and reaches a minimum at the junction between the nose and the flat portion of the plate ($x/L=0.72$), where the geometry produces a mild adverse pressure gradient. The minimum remains larger than $-0.05$, consistent with an attached laminar layer. Further downstream, $\lambda$ progressively recovers towards zero as the external pressure gradient vanishes. A similar behaviour was reported by \citet{Nagarajan2007} for an elliptical leading edge of identical aspect ratio ($AR=10$), although their computational domain extended much farther downstream. In the present configuration, the plate end corresponds only to $x/R\simeq1.39$, such that the recovery of $\lambda$ towards its asymptotic turbulent behaviour cannot yet be fully observed. In the AW case, the same overall trend is observed but with a much deeper trough ($\lambda\approx -0.1$). This stronger excursion is not caused by a larger external gradient but rather by the larger momentum thickness of the transitional/turbulent layer, which amplifies $\lambda$. The CW case shows a weaker trough ($\lambda\approx -0.06$), consistent with the smaller momentum thickness associated with wall cooling.

Panel~(b) corroborates these interpretations by showing $\theta$ and the displacement thickness $\delta^*$. From the leading edge up to $x/L\approx 0.7$, the AW case follows the same qualitative trend as the laminar solution but with significantly larger values, owing to enhanced momentum transport associated with the developing streaky fluctuations. Just upstream of the junction ($x/L\approx 0.65$) the divergence between AW and the laminar solution in $\theta$ becomes pronounced, signalling the onset of transition; beyond the junction, $\theta$ grows much more steeply, consistent with a turbulent development. The displacement thickness, more sensitive to the external pressure field, shows a clear change in slope at the junction in the laminar case; the same feature appears in AW but is attenuated, as the thicker transitional boundary layer resists the adverse gradient. In CW, both $\theta$ and $\delta^*$ remain smaller than in AW, yet they exhibit the same change of slope at the junction, suggesting that the onset of transition is not substantially shifted by wall cooling.

Panels~(c,d) provide further evidence of the transition process. The skin-friction coefficient $C_f$ in the AW case closely follows the Blasius solution up to $x/L\approx0.75$, albeit with slightly lower values, and then rises, indicating the onset of bypass transition. The levels remain below the turbulent correlation because the limited streamwise extent of the plate, discussed in §\,2.1, prevents the establishment of a fully developed turbulent regime. The CW case exhibits the same trend but with systematically larger $C_f$, reflecting the enhanced near-wall velocity gradients associated with the thinner cooled boundary layer.
The onset of transition appears at the same location as in AW, consistent with the observations of \citet{Rued1986} and \citet{Ferreira2019}, who reported no significant shift in the onset location of bypass transition due to wall cooling.
The evolution of the Stanton number in panel~(d) reinforces this picture: in CW, $St$ matches the laminar correlation for $T_g/T_w=2$ up to $x/L\approx0.25$, remains slightly above it thereafter, and then undergoes a clear departure at $x/L\approx0.75$, tending towards turbulent behaviour. This thermal signature of transition is consistent with the trends observed in $C_f$ and $\theta$. 
A final remark concerns the behaviour on the upper and lower sides of the plate: both sides exhibit the expected symmetry. 
Accordingly, in what follows, whenever a mean quantity is considered, the results from the two sides are averaged and presented together.

\begin{figure}
          \centering 
      \begin{subfigure}{0.48\textwidth}
        \includegraphics[scale=1]{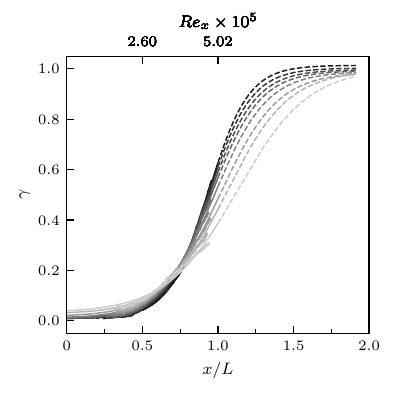}
        \caption{}
        \label{fig:intermittency_adia}
      \end{subfigure}\hfil 
      \begin{subfigure}{0.48\textwidth}
        \includegraphics[scale=1]{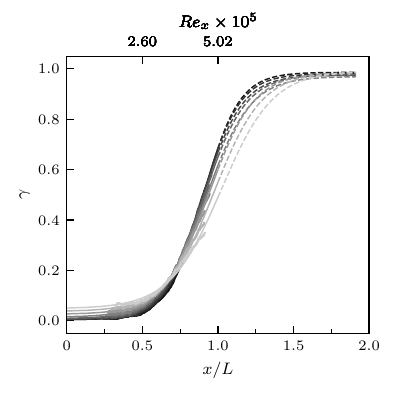}
        \caption{}
        \label{fig:intermittency_pf}
      \end{subfigure}
      \caption{Streamwise evolution of the intermittency $\gamma(x)$ for the AW (\textit{a}) and CW (\textit{b}) cases. Symbols show the intermittency computed from instantaneous $(x,z)$ slices at different wall-normal positions ($\eta=0.1$ to $0.8$, from dark to light grey). Solid lines are corresponding sigmoidal fits, extrapolated up to $x/L=2$ to illustrate the asymptotic behaviour.}

      \label{fig:intermittency}
      \end{figure}

The boundary-layer parameters thus suggest that transition, on average, initiates around $x/L \approx 0.75$. However, the snapshots discussed earlier already revealed that turbulent spots may form and convect downstream at earlier locations, leaving behind relaminarised regions. To reconcile these two perspectives, the streamwise evolution of the intermittency is now examined (figure~\ref{fig:intermittency}), providing a statistical measure of the local fraction of time during which the boundary layer is in a turbulent state. The intermittency was estimated from the instantaneous fields, by applying the detector function $D \equiv |v'|+|w'|$ proposed by \citet{Nolan2013} on $(x,z)$ slices extracted at different wall-normal locations (expressed in fractions of $\delta_{99}$) and thresholded to distinguish laminar from turbulent regions. The detector output was then averaged in the spanwise direction and over time to yield the intermittency factor $\gamma(x)$. A detailed description of the intermittency detection method is given in
Appendix~\ref{app:intermittency}, including the specific treatment near the
leading edge. Several wall-normal levels between $\eta = y / \delta_{99}=0.1$ and $0.8$ were considered, and the results are displayed together with a sigmoidal fit used to extrapolate the downstream behaviour.

In both cases, two distinct regions can be identified, separated by $x/L \approx 0.7$. Upstream of this position, the intermittency increases with wall-normal distance, indicating that turbulent activity first appears in the outer part of the boundary layer. Downstream of $x/L \approx 0.7$, the trend reverses: the intermittency is highest near the wall and decreases towards the edge. The crossing point around $x/L=0.7$ thus marks the changeover between these two regimes. A comparison between AW and CW further suggests that the transition is more gradual in AW, while it appears more abrupt under wall cooling. To quantify these observations, the streamwise positions associated with $\gamma=0.1$ and $0.5$ are summarised in table~\ref{tab:intermittency_comparison}. The beginning of transition ($\gamma=0.1$) occurs at nearly the same abscissa for AW and CW. In contrast, the midpoint of transition ($\gamma=0.5$) is systematically shifted upstream in CW by $\Delta(x/L)\approx 0.02$--$0.1$, depending on $\eta$. This suggests that, once initiated, transition develops more rapidly under wall cooling.

\begin{table}
  \centering
  \caption{Streamwise positions $x/L$ corresponding to intermittency levels
  $\gamma=0.1$ and $0.5$ for the AW and CW cases, at different wall-normal locations
  $\eta$.}
  \label{tab:intermittency_comparison}
  \begin{tabular}{c cccccccc}
    \toprule
    & \multicolumn{8}{c}{$\eta$} \\
    \cmidrule(lr){2-9}
      & 0.1 & 0.2 & 0.3 & 0.4 & 0.5 & 0.6 & 0.7 & 0.8 \\
    \midrule
    \multicolumn{9}{c}{$\gamma=0.1$ (transition beginning)} \\
    AW & 0.638 & 0.619 & 0.606 & 0.594 & 0.581 & 0.565 & 0.548 & 0.525 \\
    CW & 0.642 & 0.627 & 0.609 & 0.596 & 0.583 & 0.575 & 0.558 & 0.537 \\
    \midrule
    \multicolumn{9}{c}{$\gamma=0.5$ (transition midpoint)} \\
    AW & 0.933 & 0.937 & 0.945 & 0.956 & 0.977 & 1.014 & 1.050 & 1.113 \\
    CW & 0.906 & 0.903 & 0.904 & 0.912 & 0.927 & 0.939 & 0.968 & 1.008 \\
    \bottomrule
  \end{tabular}
\end{table}

This section has provided a global view of the flow. The FST penetrates the boundary layer, which responds through the formation of streamwise streaks. These streaks destabilise intermittently, giving rise to turbulent spots that spread and contaminate the boundary layer, confirming that the transition proceeds through the bypass route in both AW and CW. The boundary-layer parameters point to a transition onset near $x/L \approx 0.75$, whereas the intermittency measure reveals that turbulent activity is already detectable further upstream, around $x/L \approx 0.6$ at mid boundary layer height. Wall cooling has little influence on the onset of this activity but promotes a more rapid development once triggered, thereby accelerating the transition process.

  \subsection{Bypass transition mechanisms}
Having established the response of the boundary layer to the injected
turbulence, attention is now turned to the underlying physical mechanisms.
This section examines in detail the receptivity process, the streak amplification, secondary
instabilities, and spot formation processes, with emphasis on assessing the
extent to which the bypass-transition scenario is recovered in the
cooled-wall configuration.
    \subsubsection{Receptivity to free-stream turbulence \label{sec:receptivity}}
\begin{figure}
  \centering
  \begin{subfigure}{0.52\textwidth}
    \includegraphics[height=3.5cm]{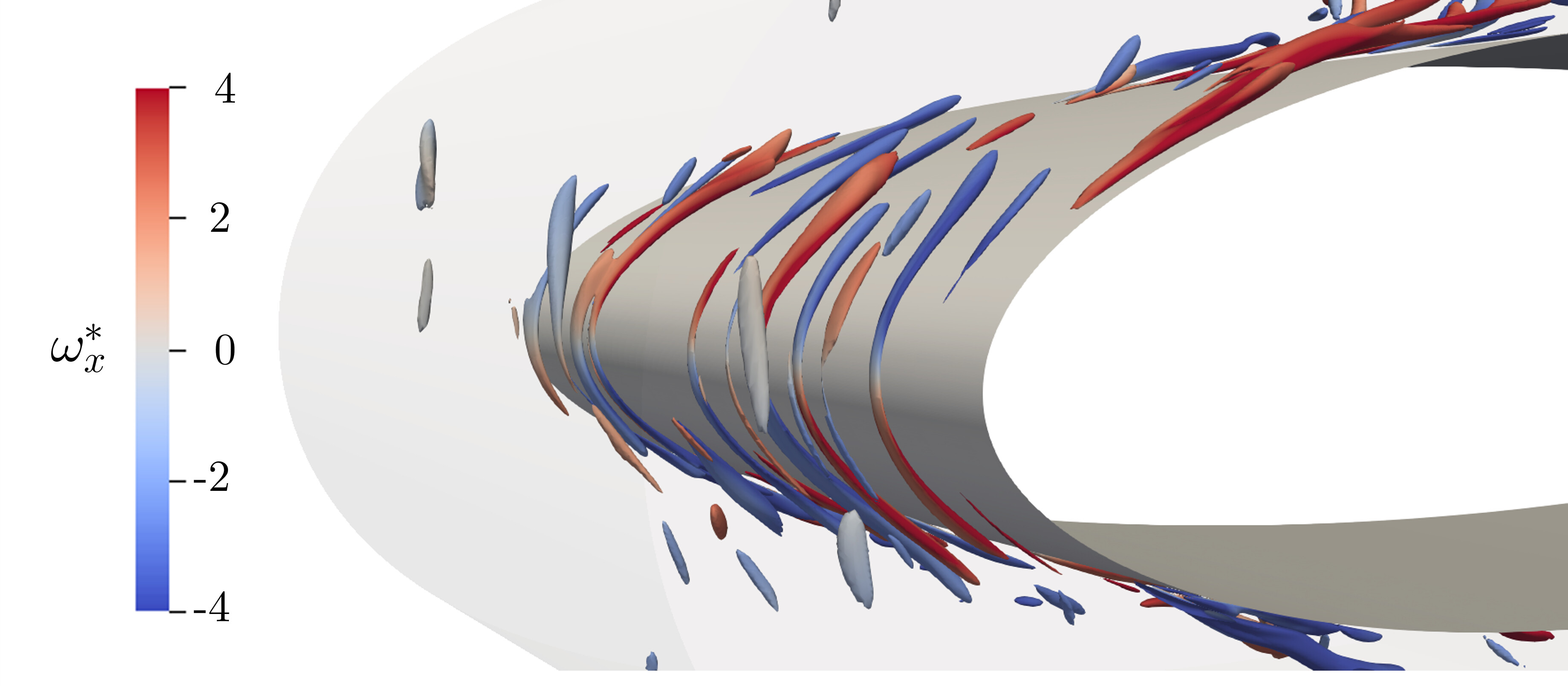}
    \caption{}
  \end{subfigure}\hfill
  \begin{subfigure}{0.46\textwidth}
    \includegraphics[height=3.5cm]{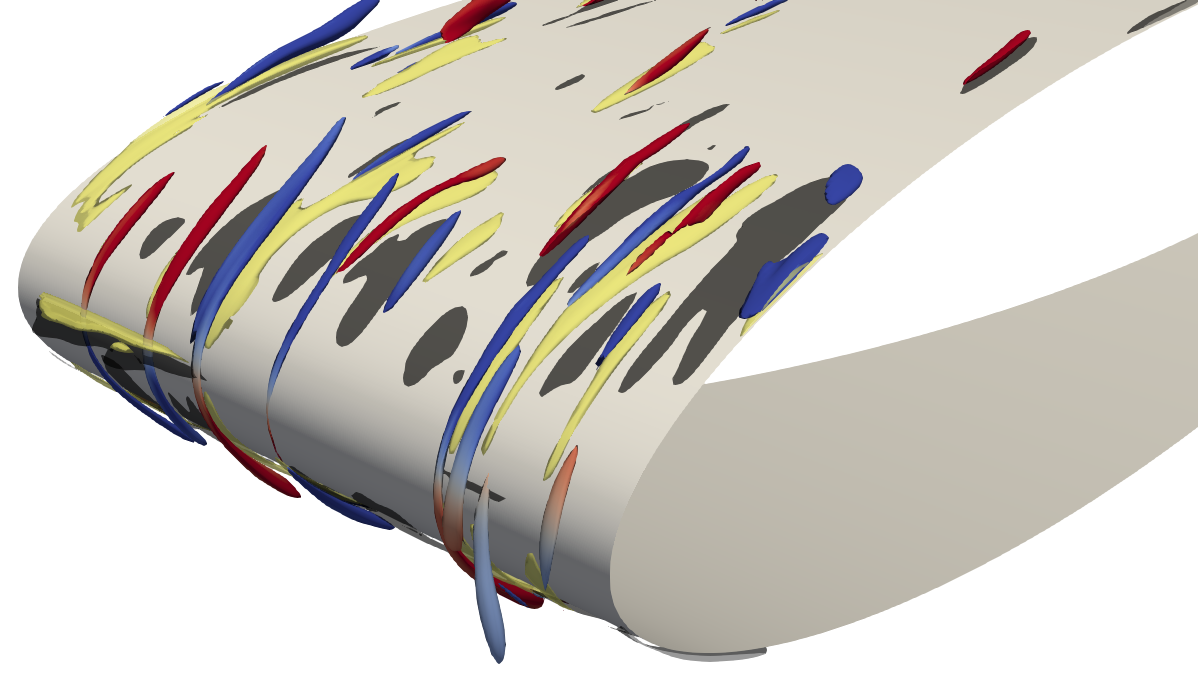}
    \caption{}
  \end{subfigure}
  \caption{Receptivity at the elliptical leading edge at $t=t_1$ (CW case). (\textit{a}) Side view: isosurfaces of the non-dimensional $Q$-criterion, $Q^\ast=3$, coloured by streamwise vorticity $\omega_x^\ast=\omega_x\, r/U_\infty$. (\textit{b}) Top-down view at the same instant: $Q^\ast=5$ (coloured by $\omega_x^\ast$) with contours of $v'_{\perp}\partial_n U_{\parallel}/(U_\infty^2/r)$ at $+0.02$ (yellow) and $-0.02$ (black).}
  \label{fig:vortex_tilting}
\end{figure}

We first examine the receptivity mechanism at the leading edge for the CW case using two instantaneous views in figure~\ref{fig:vortex_tilting}. The visualisation is restricted to the lower $20\%$ of the computational domain height,  thereby focusing on the flow structures involved in the receptivity process. The left panel provides a side view highlighting vortex wrapping around the elliptical nose. Isosurfaces of the non-dimensional $Q$-criterion are shown at $Q^\ast=3$ and coloured by the streamwise vorticity $\omega_x^\ast=\omega_x\, r/U_\infty$. As they approach the plate, incoming free-stream eddies wrap around the leading edge and stretch along the surface. Vortex tilting converts incident spanwise vorticity into streamwise vorticity of opposite sign on the upper and lower sides of the plate: an incident $\omega_y^\ast>0$ generates $\omega_x^\ast>0$ above the plate and $\omega_x^\ast<0$ below it, with the opposite arrangement for $\omega_y^\ast<0$. The present LES therefore directly resolves the leading-edge vortex-reorientation mechanism previously described by \citet{Nagarajan2007}.
The right panel shows the same instant from a more top-down perspective. Here the $Q^\ast$-isosurface level is raised to $5$ for clarity (still coloured by $\omega_x^\ast$), and contours of the non-dimensional shear-transport term $v'_{\perp}\partial_n U_{\parallel}/(U_\infty^2/r)$ are superposed (yellow for $+0.02$, black for $-0.02$). This view establishes the direct connection between the reoriented streamwise vortices and the onset of lift-up.  A vortex with $\omega_x^\ast<0$ generates negative $v'_{\perp}\partial_n U_{\parallel}$ on its left and positive values on its right, while a vortex with $\omega_x^\ast>0$ produces the opposite arrangement. These signed regions of $v'_{\perp}\partial_n U_{\parallel}$ prefigure the locations and signs of the streamwise streaks that subsequently emerge.

\begin{figure}
  \centering
  \begin{subfigure}{0.48\textwidth}
    \includegraphics[width=\linewidth]{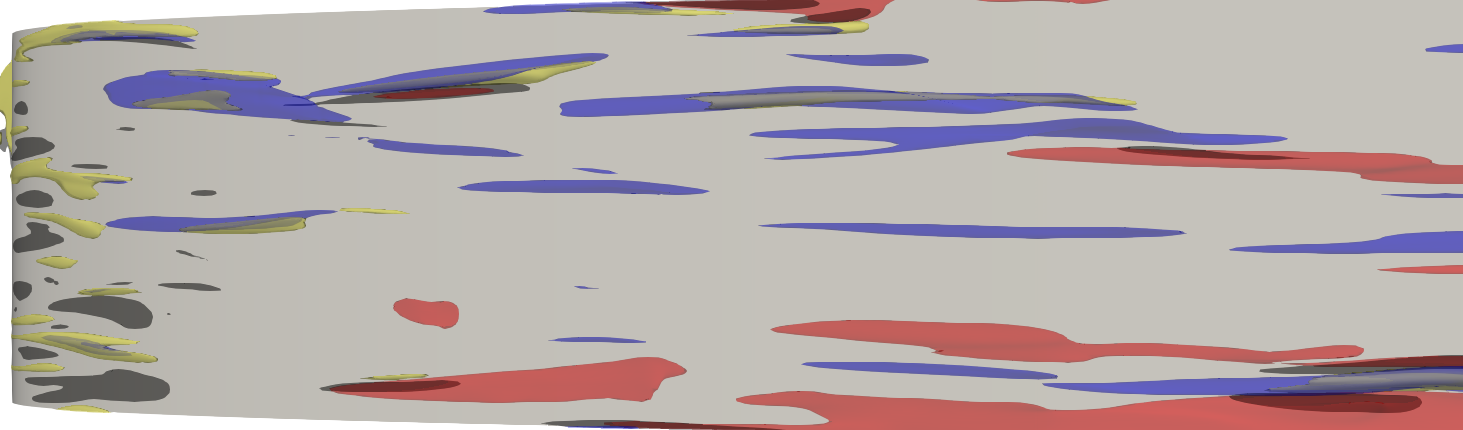}
    \caption{$u'_{\parallel}$ at $t_1$}
    \label{fig:u1}
  \end{subfigure}\hfill
  \begin{subfigure}{0.48\textwidth}
    \includegraphics[width=\linewidth]{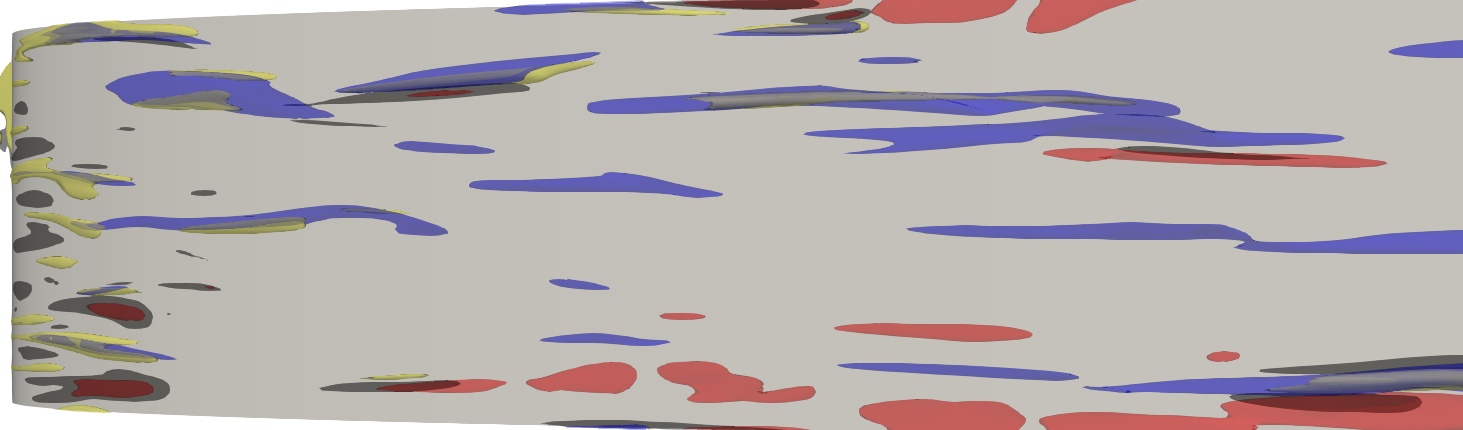}
    \caption{$T'$ at $t_1$}
    \label{fig:T1}
  \end{subfigure}

  \vspace{0.3em} 

  \begin{subfigure}{0.48\textwidth}
    \includegraphics[width=\linewidth]{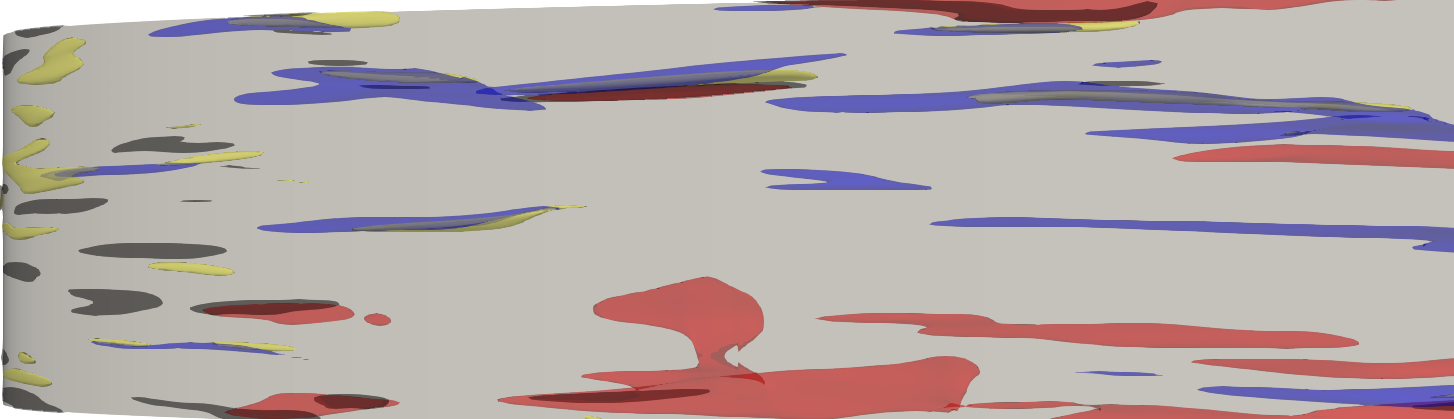}
    \caption{$u'_{\parallel}$ at $t_1+\Delta t$}
    \label{fig:u2}
  \end{subfigure}\hfill
  \begin{subfigure}{0.48\textwidth}
    \includegraphics[width=\linewidth]{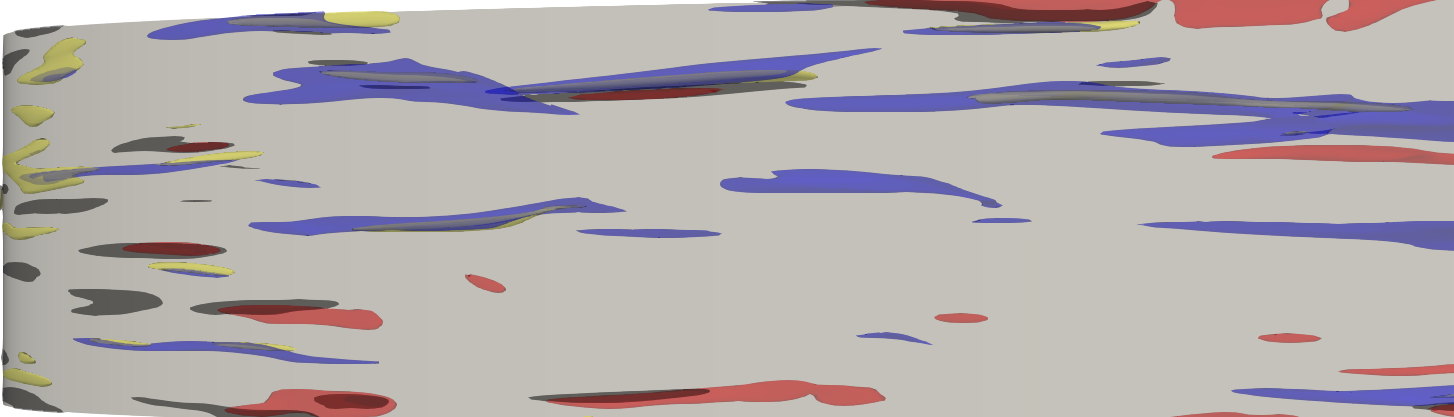}
    \caption{$T'$ at $t_1+\Delta t$}
    \label{fig:T2}
  \end{subfigure}

\caption{Plan view of streak formation (CW case). All panels show positive/negative contours of $v'_{\perp}\partial_n U_{\parallel}/(U_\infty^2/r)$. (\textit{a,c})Superposed contours of $u'_{\parallel}=\pm 0.12\,U_\infty$ at $t=t_1$ and $t=t_1+\Delta t$; (\textit{b,d}) superposed contours of $T'=\pm 0.06\,T_\infty$ at the same instants.}
  \label{fig:liftup}
\end{figure}

The plan view shown in figure~\ref{fig:liftup} further clarifies how the shear-transport field precedes and structures the streaks. All panels display positive/negative contours of $v'_{\perp}\partial_n U_{\parallel}/(U_\infty^2/r)$; panels~(a,c) superpose streamwise velocity fluctuations $u'_{\parallel}=\pm 0.12\,U_\infty$ (blue negative, red positive), while panels~(b,d) superpose temperature fluctuations $T'=\pm 0.06\,T_\infty$ (blue negative, red positive). At $t=t_1$ [panels~(a,b)], patches of $v'_{\perp}\partial_n U_{\parallel}$ are organised in streamwise structures. Positive $v'_{\perp}\partial_n U_{\parallel}$ subsequently gives rise to low-velocity streaks (negative $u'_{\parallel}$), through the upward displacement of slower near-wall fluid, while negative $v'_{\perp}\partial_n U_{\parallel}$ precedes high-velocity streaks (positive $u'_{\parallel}$), through the downward displacement of faster outer fluid. This causal ordering is confirmed at $t=t_1+\Delta t$ [panels~(c,d)]: the $v'_{\perp}\partial_n U_{\parallel}$ field appears first, then $u'_{\parallel}$ streaks develop and persist downstream even after the local $v'_{\perp}\partial_n U_{\parallel}$ signature weakens. The CW simulation further demonstrates that the same mechanism organises the temperature fluctuations: thermal streaks are collocated with the streamwise velocity streaks and inherit their sign through wall-normal advection (positive $v'_{\perp}\partial_n U_{\parallel}$ brings cooler fluid upward, producing negative $T'$, and vice versa). This result extends the velocity--thermal streak organisation predicted by OPT \citep{Tumin2003, Vermeersch2009Thesis} to a LES capturing bypass transition triggered by FST under wall cooling.
The amplitude of the thermal streaks is comparable to their velocity counterparts. Altogether, this temporal sequence establishes the shear-transport field as a precursor of streak formation: shear-transport patches seed and organise the ensuing velocity streaks, while wall cooling produces a thermal counterpart governed by the same lift-up dynamics.

\begin{figure}
              \centering
              \begin{subfigure}{\linewidth}
                \centering
                \includegraphics[scale=1]{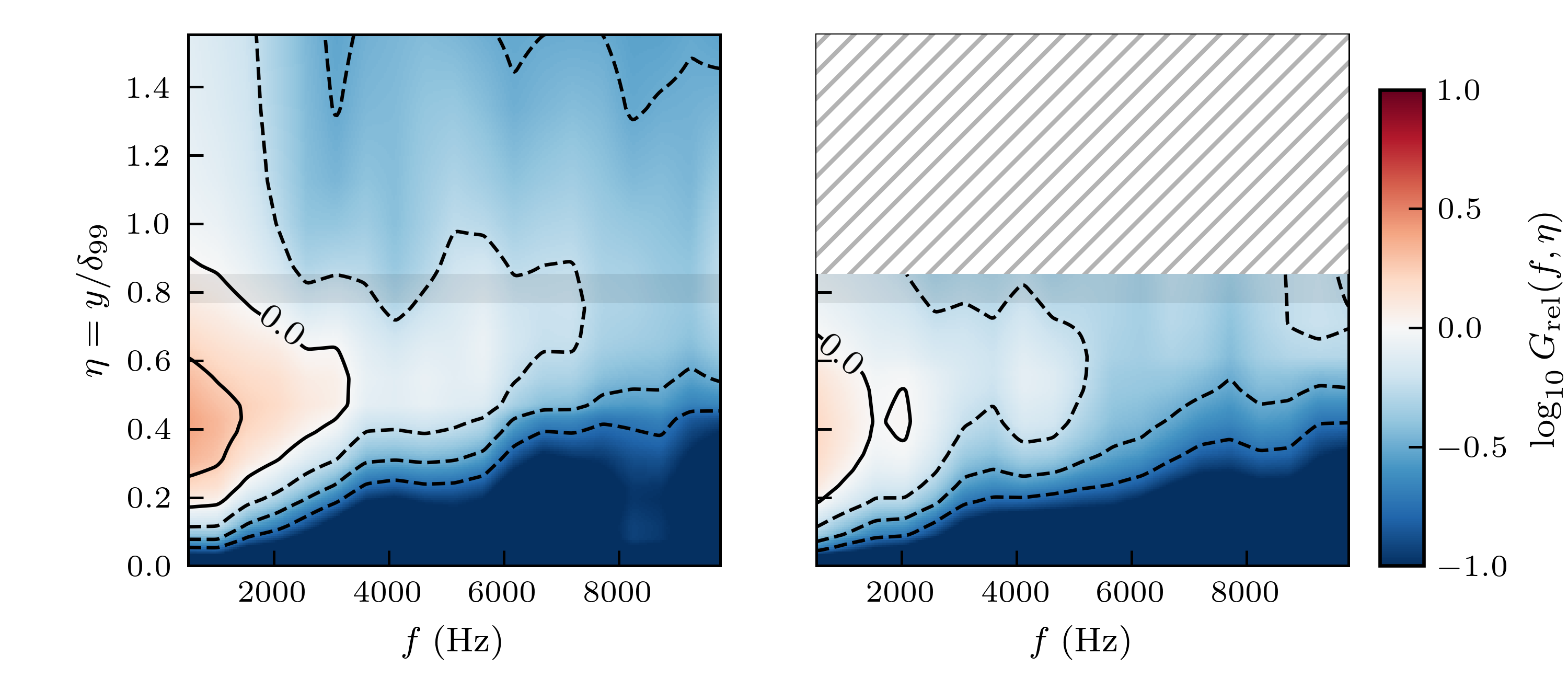}
                \caption{$x/L=0.07$}
              \end{subfigure}\\[0.3cm]
              \begin{subfigure}{\linewidth}
                \centering
                \includegraphics[scale=1]{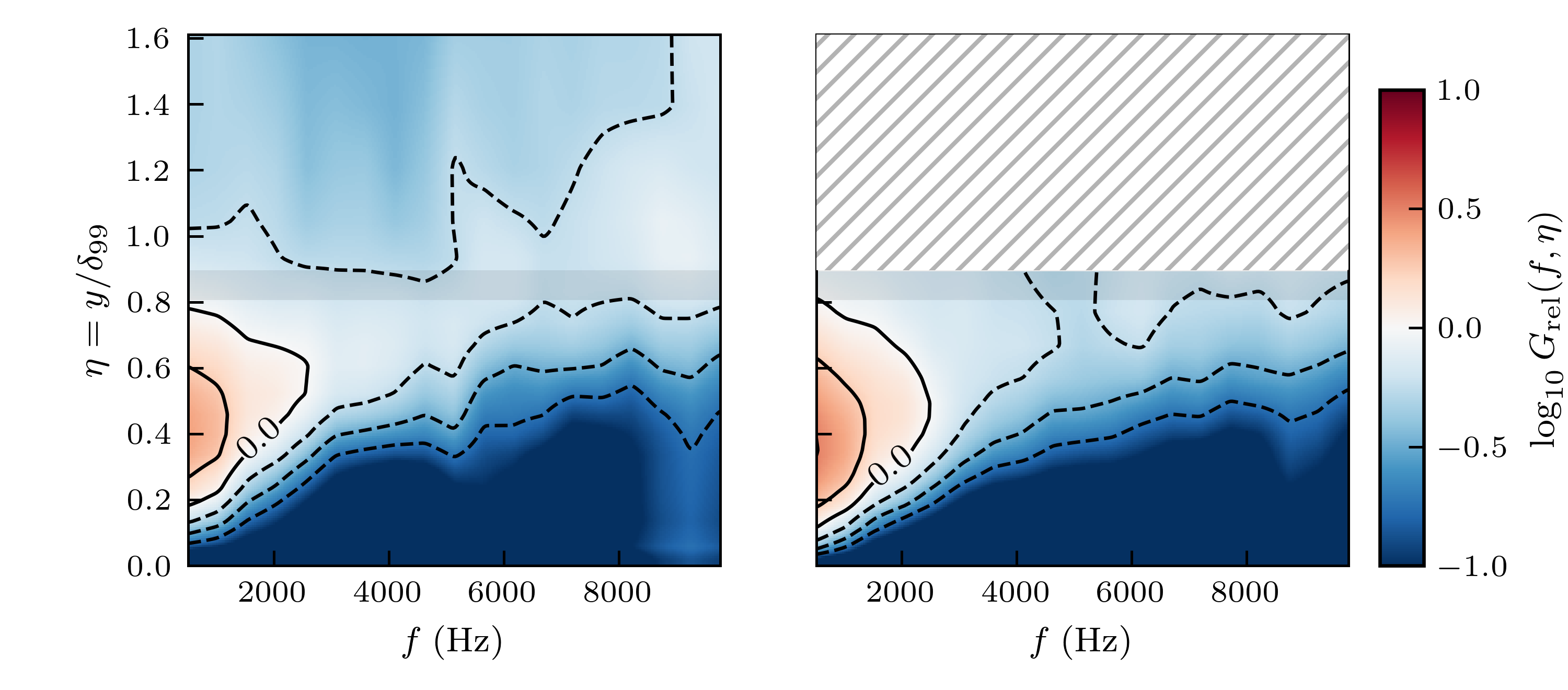}
                \caption{$x/L=0.24$}
              \end{subfigure}\\[0.3cm]
              \begin{subfigure}{\linewidth}
                \centering
                \includegraphics[scale=1]{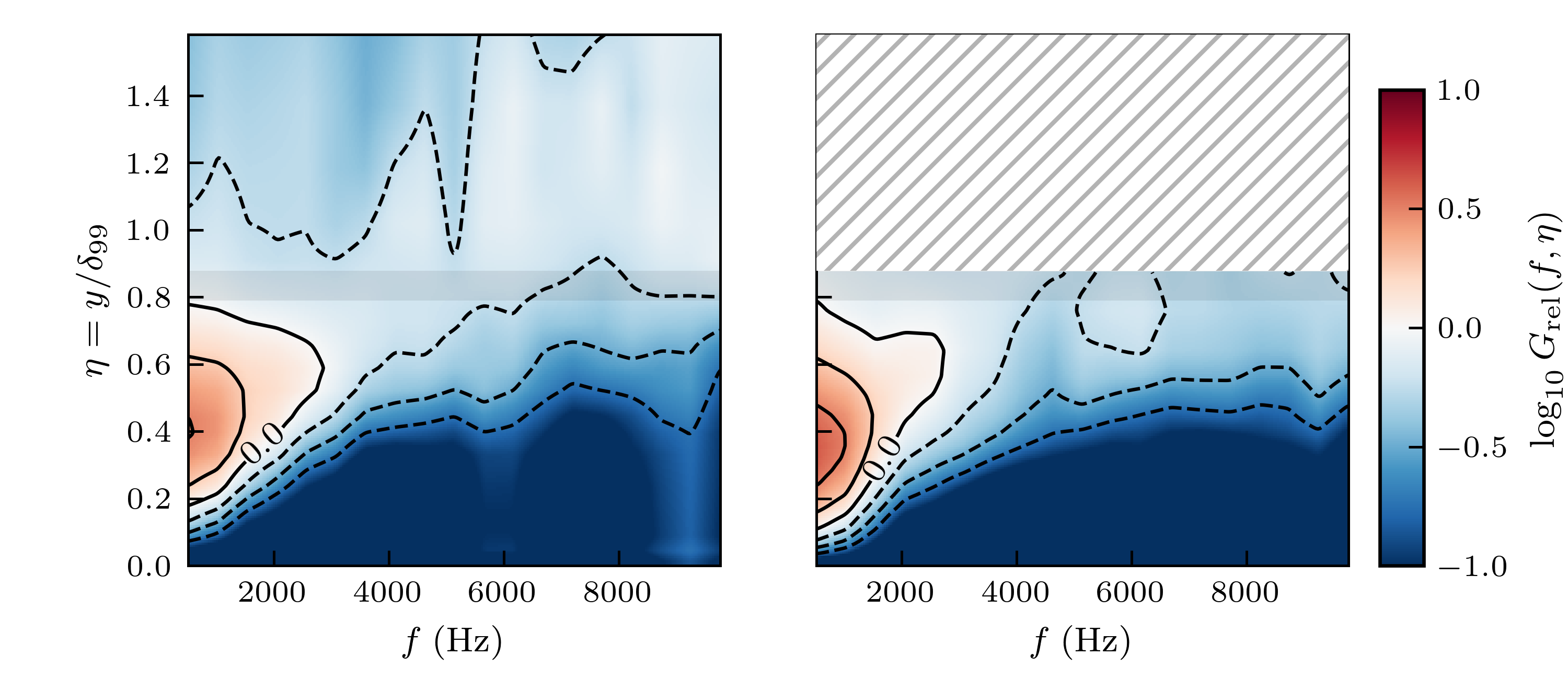}
                \caption{$x/L=0.38$}
              \end{subfigure}
\caption{Maps of the relative gain 
$\log_{10} G_{\mathrm{rel}}(f,\eta)$ 
of the streamwise velocity for the AW (left) 
and CW (right) cases at a given streamwise station. 
Translucent
grey band: reference used to normalised spectra, defined as the average
over the outermost 10\% of the available wall-normal domain of the CW case; 
hatched area: unsampled region in the CW configuration.}
              \label{fig:gain_receptivity}
            \end{figure}

Having established qualitatively that wrapped vortices generate patches of 
$v'_{\perp}\partial_n U_{\parallel}$ which seed velocity and thermal streaks, we now quantify how 
free-stream perturbations penetrate the boundary layer by analysing shear-sheltering.
Figure~\ref{fig:gain_receptivity} shows maps of the relative gain for the 
AW (left) and the CW (right) at three streamwise 
stations: (a) $x/L=0.07$, (b) $x/L=0.24$ and (c) $x/L=0.38$.
We define the relative gain of the streamwise velocity as
\[
G_{\mathrm{rel}}(f,\eta)
\;=\;
\frac{E_{u}(f,\eta)}
     {\big\langle E_{\mathrm{tot}}(f,\eta)\big\rangle_{\eta\in[\eta_{\mathrm{low}},\,\eta_{\mathrm{top}}]}}\!,
\qquad 
\eta=y/\delta_{99},
\]
where \(E_{u}(f,\eta)\) is the premultiplied power spectral density of the
streamwise fluctuations \(u'_{\parallel}\). The reference spectrum in the denominator is the
near-edge average of the total premultiplied energy,
\[
E_{\mathrm{tot}}(f,\eta)
\,=\, E_{u}(f,\eta)+E_{v}(f,\eta)+E_{w}(f,\eta).
\]
The power spectra are estimated using Welch's method. To ensure sufficiently
converged spectral estimates while retaining adequate frequency resolution,
the segment length is limited to 5000 samples. With a sampling frequency of
$f_s=1/\Delta t=2.57\times10^6\,\mathrm{Hz}$, this yields a frequency
resolution of $\Delta f=f_s/N_{\mathrm{perseg}}\approx514\,\mathrm{Hz}$.
Consequently, although the analysis formally considers frequencies above
$182\,\mathrm{Hz}$, the lowest non-zero resolved frequency is approximately
$514\,\mathrm{Hz}$.
The wall-normal probe distributions are not identical in the AW and CW configurations.
Consequently, the accessible range in the non-dimensional coordinate $\eta$ differs between
the two cases. To enable a like-for-like comparison, the near-edge reference interval 
$[\eta_{\mathrm{low}},\eta_{\mathrm{top}}]$ is chosen from the CW case as the 
outermost 10\% of its available wall-normal extent in $\eta$, and the same
$\eta$ interval is used to normalise both panels. The interval
$[\eta_{\mathrm{low}},\eta_{\mathrm{top}}]$ is highlighted by a translucent
grey band. The unsampled 
portion above the highest probe in the CW case appears hatched.

This normalisation reveals the frequency ranges and wall-normal regions where the
boundary-layer dynamics either amplify \((\log_{10}G_{\mathrm{rel}}>0)\) or attenuate
\((\log_{10}G_{\mathrm{rel}}<0)\) the external fluctuations that have penetrated the
boundary layer (as sampled in the near-edge reference band, \(\eta\approx 0.80\text{--}0.85\)).
Shear-sheltering is clearly evident: the boundary layer filters high-frequency content
by attenuating it. Tracking the \(\log_{10}G_{\mathrm{rel}}=-0.5\) isocontour, we observe that at
\(x/L=0.07\) the highest frequencies \((f\gtrsim 7500\,\mathrm{Hz})\) are preferentially damped
within the lower half of the layer \((\eta\le 0.5)\). Further downstream the effect strengthens:
by \(x/L=0.38\) frequencies up to \(\approx 4500\,\mathrm{Hz}\) are attenuated to \(-0.5\) over
\(\eta\le 0.5\). The mechanism is essentially the same in the AW and CW configurations, notably
in terms of penetration depth. 
In contrast, the lowest resolved frequencies \((f\lesssim 3000\,\mathrm{Hz})\)
are amplified within the boundary layer, with a pronounced peak near \(\eta\approx 0.5\) at the lowest resolved frequencies.
More precisely, at \(x/L=0.38\) the maximum occurs around \(\eta\approx 0.45\) for AW and slightly
lower, \(\eta\approx 0.40\), for CW. 
These wall-normal locations correspond to the positions at which the streak-amplitude profiles attain their maximum. 

These maps quantify the shear-sheltering process: high-frequency disturbances are progressively excluded from the inner boundary layer, whereas low-frequency disturbances penetrate deeper and undergo strong amplification. The coincidence between the low-frequency amplification maximum and the wall-normal location of the velocity streaks directly links the frequency-selective penetration of free-stream disturbances to subsequent streak amplification.
The comparison between AW and CW further demonstrates that wall cooling does not alter the shear-sheltering mechanism: the frequency-dependent penetration of free-stream disturbances remains essentially unchanged, while the low-frequency amplification maximum is displaced slightly towards the wall in CW. 
Together with the vortex-wrapping and lift-up mechanisms illustrated in figures~\ref{fig:vortex_tilting} and \ref{fig:liftup}, these results establish that the complete leading-edge receptivity sequence (vortex reorientation, shear sheltering, lift-up and streak formation) is preserved under strong wall cooling.

    \subsubsection{Streak development and growth}

The receptivity stage and the initial development of the streaks appear broadly similar in the AW and CW configurations. The main differences are the emergence, in the CW case, of thermal streaks correlated with the velocity streaks, and a slightly lower wall-normal position of the streamwise velocity fluctuations maximum compared with AW. We now characterise more precisely the structure and growth of these fluctuations in the pre-transitional region, where they take the form of coherent streaks.

\begin{figure}
          \centerline{\includegraphics[scale=1]{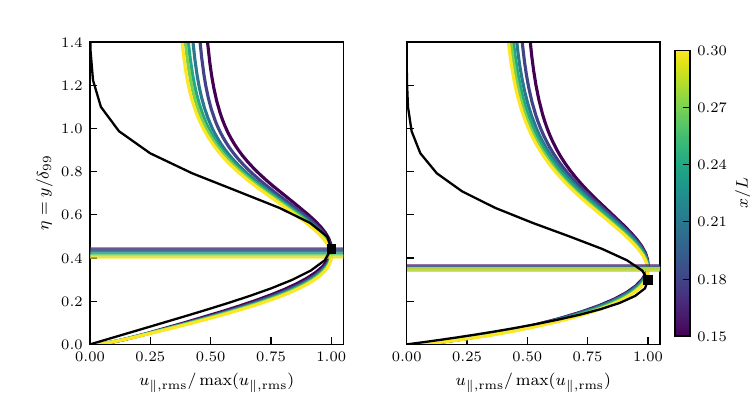}}
          \caption{Wall-normal profiles of the streamwise fluctuation intensity
          $u_{\parallel,\mathrm{rms}}$ at selected streamwise locations $x/L$. Left: AW;
          right: CW. Profiles are normalised by their local maximum, and the corresponding
          wall-normal position is indicated by a faint horizontal line. The solid black curve
          in each panel shows the OPT prediction for the corresponding
          wall condition; the square marks its maximum.}
          \label{fig:TPO_streaks}
        \end{figure}

Figure~\ref{fig:TPO_streaks} compares wall-normal profiles of the streamwise fluctuation intensity with OPT predictions for both wall-thermal
conditions. The quantity considered is the root-mean-square of the streamwise velocity fluctuation in the plate-aligned frame,
$u_{\parallel,\mathrm{rms}} = \sqrt{\langle u_{\parallel}'^{\,2} \rangle}$,
which provides a robust measure of the streak amplitude in the pre-transitional region. Profiles are extracted at several streamwise locations and normalised by their local maximum; the wall-normal position corresponding to this maximum is indicated by a faint horizontal line. The stations are chosen in the interval $0.15\le x/L\le 0.30$, where the streaks are already formed yet not destabilised.
For the AW case, the LES profiles recover the OPT prediction, with a rapid near-wall increase, a single outer maximum, and a gradual decay towards the boundary-layer edge. The OPT envelope vanishes near the edge since it represents the streak contribution only and does not account for FST. The peak is located at $\eta\simeq 0.43$, essentially reproducing the OPT prediction of $\eta=0.44$. In the CW case, the profiles retain the same overall shape, while the maximum shifts towards the wall to $\eta\simeq 0.37$. The LES therefore confirms the wallward displacement predicted by OPT, although the displacement is weaker than predicted ($\eta=0.30$ in OPT). Overall, wall cooling preserves the characteristic wall-normal structure of the streaks while shifting their maximum towards the wall.

\begin{figure}
          \centerline{\includegraphics[scale=1]{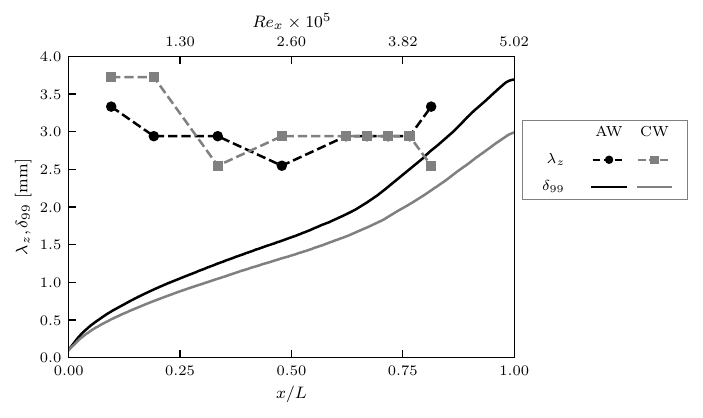}}
\caption{
Downstream evolution of the boundary-layer thickness \(\delta_{99}(x)\) 
and the spanwise streak spacing \(\lambda_z(x)\)  on the upper surface, 
for the AW and the CW case. }
          \label{fig:spacing_streaks}
        \end{figure}

Having established the wall-normal structure of the streaks and its consistency with OPT, we now examine their spanwise scale. 
The lateral organisation is quantified by extracting a spacing \(\lambda_z(x)\) from the two-point correlation of the streamwise fluctuations. 
At each streamwise station \(x\), a rake of \(N\approx80\) spanwise probes provides time series \(u(z,t), v(z,t)\), from which \(u'_{\parallel}(z,t)\) is formed. 
The normalised correlation
\[
R_{uu}(\Delta z)=\frac{\langle u'_{\parallel}(z,t)\,u'_{\parallel}(z+\Delta z,t)\rangle_t}{\sigma(z)\,\sigma(z+\Delta z)},
\]
where $\sigma(z)$ is the standard deviation over time of $u'_{\parallel}$, is computed for all discrete shifts using periodic wrapping in \(z\), and recentered so that \(\Delta z=0\) lies at mid-span. 
The spacing \(\lambda_z\) is defined as the distance between the first pair of minima of \(R_{uu}\) immediately to the left and right of \(\Delta z=0\). 
Figure~\ref{fig:spacing_streaks} juxtaposes the resulting \(\lambda_z(x)\) with the concurrent boundary-layer thickness \(\delta_{99}(x)\), enabling a like-for-like comparison between the outer growth of the layer and the spanwise arrangement of the streaks.
Over the range considered, the spanwise scale remains fairly constant, with \(\lambda_z\) lying between \(2.5\) and \(3.5\,\mathrm{mm}\). 
The CW configuration exhibits a larger variability, starting from slightly higher values upstream and dipping somewhat lower downstream, yet both cases tend towards \(\lambda_z \approx 3\,\mathrm{mm}\) further downstream. 
This magnitude is comparable to the local \(\delta_{99}\), indicating that \(\lambda_z/\delta_{99}\) tends to unity. Overall, wall cooling leaves the characteristic spanwise scale of the velocity streaks essentially unchanged.

\begin{figure}
    \centering

    \begin{subfigure}[t]{0.315\textwidth}
        \centering
        \includegraphics[scale=1]{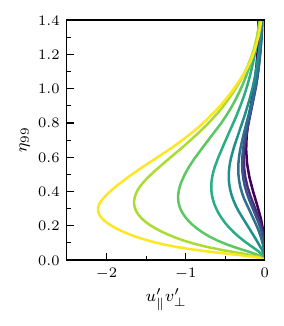}
        \caption{}
    \end{subfigure}
    \hfill
    \begin{subfigure}[t]{0.24\textwidth}
        \centering
        \includegraphics[scale=1]{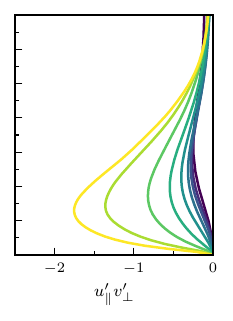}
        \caption{}
    \end{subfigure}
    \hfill
    \begin{subfigure}[t]{0.38\textwidth}
        \centering
        \includegraphics[scale=1]{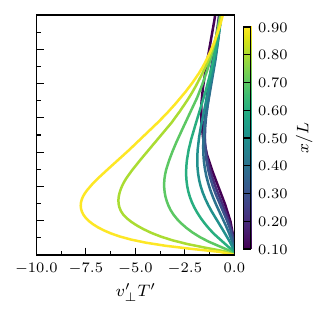}
        \caption{}
    \end{subfigure}

    \caption{
    Wall-normal profiles of the Reynolds shear stress $u'_{\parallel}v'_{\perp}$ for the AW (\textit{a}) and CW (\textit{b}) cases, and of the turbulent heat flux $v'_{\perp}T'$ for the CW case (\textit{c}). Profiles are shown at selected streamwise locations $x/L$ and are coloured by streamwise position.
    }
    \label{fig:uv_vT_profiles}
\end{figure}

We next examine whether wall cooling similarly preserves the streak dynamics by considering the fluctuation correlations associated with momentum and heat transport. Figure~\ref{fig:uv_vT_profiles} compares the evolution of $u'_{\parallel}v'_{\perp}$ for the AW and CW configurations, together with the corresponding $v'_{\perp}T'$ profiles in the CW case. Profiles are averaged between the upper and lower sides of the plate. In both thermal conditions, the upstream region ($x/L \lesssim 0.4$) is characterised by very weak values of $u'_{\parallel}v'_{\perp}$, consistent with a laminar boundary layer. Although its magnitude remains small, the profile already exhibits a lobe-like structure with a minimum located in the same wall-normal region as the streak maximum, reflecting the wall-normal transport of streamwise momentum associated with the lift-up mechanism. Further downstream, a well-defined negative lobe emerges around $x/L \simeq 0.5$, marking the onset of significant amplification of wall-normal momentum transport. As the flow develops, this lobe intensifies and remains centred in the region where the streak amplitude reaches its maximum, indicating that the dominant transport mechanism is closely tied to the streak dynamics.
The comparison between AW and CW shows that the structure and wall-normal location of the $u'_{\parallel}v'_{\perp}$ profiles remain very similar, with a slightly lower position of the minimum in the CW case, consistent with the downward shift of the streak core observed previously. This close agreement indicates that the dynamical mechanism underlying bypass transition is not fundamentally altered by wall cooling. However, at a given streamwise location, the magnitude of $|u'_{\parallel}v'_{\perp}|$ is systematically lower in the CW case. This apparent discrepancy with the earlier observation of a comparable or even accelerated transition in terms of intermittency highlights that amplitude-based and intermittency-based indicators capture different aspects of the transition process. While $u'_{\parallel}v'_{\perp}$ reflects the intensity of the momentum transport associated with coherent structures, intermittency quantifies the spatial and temporal extent of turbulent regions. In addition, the reduced amplitudes observed in the CW case may be influenced by density variations and by the use of unweighted fluctuations.
The turbulent heat flux $v'_{\perp}T'$ in the CW configuration exhibits a similar wall-normal distribution to that of $u'_{\parallel}v'_{\perp}$, with the appearance of a negative lobe that strengthens downstream. This behaviour reflects the transport of colder fluid away from the wall and demonstrates that the development of thermal fluctuations is directly coupled to the velocity streaks. The close correspondence between $u'_{\parallel}v'_{\perp}$ and $v'_{\perp}T'$ indicates that the bypass transition process simultaneously establishes momentum and heat transport through the same underlying mechanisms, with thermal streaks behaving as the thermal counterpart of the velocity streaks, exhibiting a similar wall-normal structure and downstream evolution.

\subsubsection{Conditional averaging of high- and low-velocity streaks}
While traditional turbulence statistics based on root-mean-square fluctuations provide valuable information on the overall disturbance levels, they do not distinguish between positive and negative streamwise velocity perturbations and therefore cannot separately characterise the contributions of high- and low-velocity streaks \citep{Hernon2007,Nolan2013}. To address this limitation, \citet{Hernon2007} examined the maximum positive and negative fluctuation amplitudes separately, revealing distinct behaviours of high- and low-velocity streaks during bypass transition. As discussed by \citet{Nolan2013}, however, focusing solely on extreme values discards much of the available information and may lead to estimates that are unduly influenced by a few exceptionally strong events. 

In the present work, a conditional averaging procedure is adopted to separate the respective contributions of high- and low-velocity streaks while retaining information from the complete set of instantaneous three-dimensional flow fields.
Each point $(x,y,z)$ is classified according to the sign of the instantaneous streamwise velocity fluctuation $u'_{\parallel}$. High-velocity (HV) and low-velocity (LV) events are defined by the conditions $u'_{\parallel}>\varepsilon$ and $u'_{\parallel}<-\varepsilon$, respectively, where $\varepsilon = p_c U_\infty/100$ denotes a threshold expressed as a fraction of the free-stream velocity. Two values are considered in the following, namely $p_c=0$ and $p_c=12$, allowing us to probe both the full fluctuation field and the most energetic structures. Conditional averages are then constructed separately over HV and LV events, and subsequently averaged in the spanwise direction to obtain two-dimensional statistics. In the pre-transitional region, these structures correspond to coherent streaks generated by the lift-up mechanism, so that the present conditioning directly separates high- and low-velocity streaks. Further downstream, as the flow undergoes breakdown, the same procedure continues to distinguish regions of positive and negative streamwise fluctuations, although the underlying structures are no longer strictly streak-like.
Since the present approach relies on a finite number of snapshots, its statistical convergence must be assessed. To this end, the quantities extracted from the instantaneous fields are compared to reference values obtained from time-averaged turbulence statistics over the entire simulation. This validation, reported in Appendix~\ref{app:validation_MC_snapshots} for both the upper and lower side of the plate, demonstrates that the present sampling is sufficient to accurately capture both the amplitude and wall-normal position of the fluctuations.

\begin{figure}
          \centering 
      \begin{subfigure}{0.49\textwidth}
        \includegraphics[scale=1]{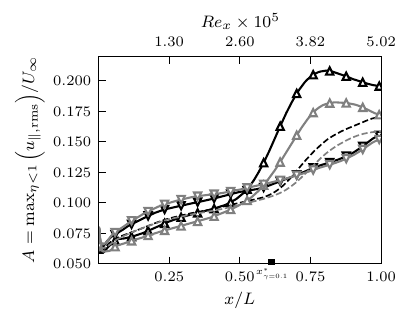}
        \caption{}
        \label{fig:amplitude_MC_0pc}
      \end{subfigure}\hfil 
      \begin{subfigure}{0.49\textwidth}
        \includegraphics[scale=1]{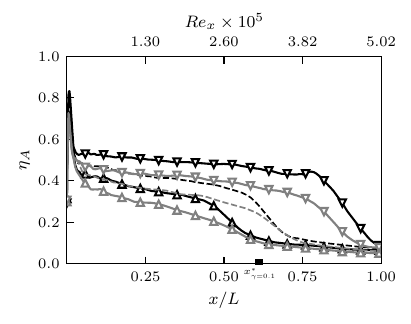}
        \caption{}
        \label{fig:position_MC_0pc}
      \end{subfigure}

\vspace{0.5em}

\begin{center}
{\footnotesize
\setlength{\fboxsep}{4pt}
\setlength{\fboxrule}{0.3pt}

\fcolorbox{black!50}{white}{%
\begin{tabular}{
    c
    @{\hspace{1.1em}} c
    @{\hspace{1.4em}} c
    @{\hspace{1.4em}} c
}

& Non-cond. & HV & LV \\[0.1em]

AW
&
\raisebox{0.3ex}{%
\tikz{\draw[line width=1.2pt,
            dash pattern=on 4pt off 2pt]
            (0,0) -- (0.65,0);}
}
&
\raisebox{0.3ex}{%
\tikz{
    \draw[line width=1.2pt] (0,0) -- (0.65,0);
    \node[draw,
          line width=1.2pt,
          regular polygon,
          regular polygon sides=3,
          inner sep=1.1pt,
          fill=white]
          at (0.325,0) {};
}}
&
\raisebox{0.3ex}{%
\tikz{
    \draw[line width=1.2pt] (0,0) -- (0.65,0);
    \node[draw,
          line width=1.2pt,
          regular polygon,
          regular polygon sides=3,
          shape border rotate=180,
          inner sep=1.1pt,
          fill=white]
          at (0.325,0) {};
}}
\\[0.15em]

CW
&
\raisebox{0.3ex}{%
\tikz{\draw[gray,
            line width=1.2pt,
            dash pattern=on 4pt off 2pt]
            (0,0) -- (0.65,0);}
}
&
\raisebox{0.3ex}{%
\tikz{
    \draw[gray,line width=1.2pt] (0,0) -- (0.65,0);
    \node[draw=gray,
          line width=1.2pt,
          regular polygon,
          regular polygon sides=3,
          inner sep=1.1pt,
          fill=white]
          at (0.325,0) {};
}}
&
\raisebox{0.3ex}{%
\tikz{
    \draw[gray,line width=1.2pt] (0,0) -- (0.65,0);
    \node[draw=gray,
          line width=1.2pt,
          regular polygon,
          regular polygon sides=3,
          shape border rotate=180,
          inner sep=1.1pt,
          fill=white]
          at (0.325,0) {};
}}

\end{tabular}%
}}
\end{center}

\vspace{0.2em}

      \caption{
      Downstream evolution of the streak amplitude $A=\max_{\eta<1}(u_{\parallel,\mathrm{rms}})/U_\infty$ (\textit{a}) and corresponding wall-normal position $\eta_A$ (\textit{b}) for the conditional analysis with $p_c=0$. Results are shown for the AW and CW configurations, for non-conditioned statistics and conditional HV and LV events.
      }
      \label{fig:MC_0pc}
      \end{figure}

The results of the conditional analysis for $p_c=0$ are presented in figure~\ref{fig:MC_0pc}. The streamwise evolution of the fluctuation amplitude $A=\max_{\eta<1}(u_{\parallel,\mathrm{rms}})/U_\infty$ is shown in the left panel, while the corresponding wall-normal position $\eta_A$ is reported in the right panel. Results on the upper and lower side are averaged. The adiabatic (AW) and cold-wall (CW) cases are shown in black and grey, respectively. Dashed lines correspond to the non-conditioned statistics, while solid lines denote the conditional averages associated with high-velocity (HV, $\triangle$) and low-velocity (LV, $\triangledown$) events. 
The streamwise interval where both cases exhibit an intermittency level $\gamma(\eta=0.5)>0.1$ is highlighted by a black rectangle on the $x$-axis, labelled $x^{\ast}_{\gamma=0.1}$, where $x^\ast=x/L$.
The same conventions are used in both panels.

Both AW and CW cases exhibit the formation of streamwise streaks with an initial amplitude of approximately $6\%$ of the free-stream velocity, irrespective of their sign. The receptivity region preceding the onset of streak growth is very short, extending over less than $5\%$ of the plate length. In the sense of \citet{Fransson2005}, this region corresponds to the initial adjustment of free-stream disturbances to the boundary-layer length scales, characterised by a slower growth of $u_{\mathrm{rms}}$ prior to the linear amplification of streaks. Beyond this region, the streak amplitude increases steadily up to $x/L\simeq 0.5$.
In the pre-transitional regime, the non-conditioned statistics confirm that wall cooling has little influence on streak amplitudes, thereby extending the OPT results of \citet{Vermeersch2009Thesis} to bypass transition triggered by FST under wall cooling. Conditional averaging reveals a subtle redistribution between positive and negative velocity fluctuations: in the CW case, LV streaks are somewhat more energetic, whereas HV streaks exhibit slightly reduced amplitudes compared to the AW configuration. In both AW and CW cases, LV streaks exhibit larger amplitudes than HV streaks and are located farther from the wall. This asymmetric organisation confirms the DNS observations of \citet{Nolan2013}, who reported LV streaks residing higher within the boundary layer than HV streaks. The wall-normal position of the maximum streak amplitude also exhibits a slight downstream decrease for both HV and LV structures over the pre-transitional region ($x/L<0.5$).
A marked change in behaviour is observed beyond $x/L\simeq 0.5$. The amplitude of HV structures undergoes a rapid increase, accompanied by a pronounced shift of their maximum towards the wall. This occurs upstream of the location $x^{\ast}_{\gamma=0.1}$. In fact, $x^{\ast}_{\gamma=0.1}$ corresponds to the location at which the wall-normal migration of the HV structures saturates.
In contrast, LV structures display a much smoother evolution. Their amplitude continues to grow gradually, and the wall-normal position of their maximum remains nearly unchanged throughout this region. Downstream of $x^{\ast}_{\gamma=0.1}$, the growth of HV structures persists, reaching amplitudes of about $20\%$ of the free-stream velocity in the AW case and $17.5\%$ in the CW case. At the same time, their maximum is located very close to the wall, around $\eta\simeq 0.08$. The peak amplification of HV structures is attained near $x/L\simeq 0.75$, which coincides with the streamwise location where transition signatures appear in boundary-layer quantities such as the skin-friction coefficient $C_f$. At this location, the LV structures also undergo a sudden inward shift, indicating that the transition process is fully established. A similar downward migration of low-velocity streaks after the onset of transition was also reported by \citet{Nolan2013}.

\begin{figure}
          \centering 
      \begin{subfigure}{0.49\textwidth}
        \includegraphics[scale=1]{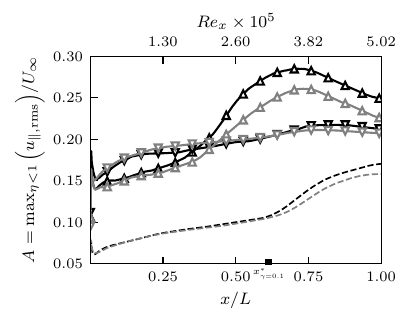}
        \caption{}
        \label{fig:amplitude_MC_12pc}
      \end{subfigure}\hfil 
      \begin{subfigure}{0.49\textwidth}
        \includegraphics[scale=1]{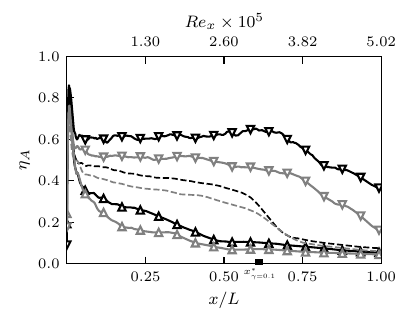}
        \caption{}
        \label{fig:position_MC_12pc}
      \end{subfigure}
      \caption{Same as figure~\ref{fig:MC_0pc}, but for a conditional threshold $p_c=12$.}
      \label{fig:MC_12pc}
      \end{figure}

To further isolate the most energetic structures, figure~\ref{fig:MC_12pc} presents the results of the conditional analysis using a threshold of $p_c=12$. The non-conditioned statistics are retained for reference. In the upstream region ($x/L<0.5$), LV streaks remain more energetic than HV streaks. However, this trend reverses at $x/L\simeq 0.36$ for AW and $x/L\simeq 0.44$ for CW. Compared to the $p_c=0$ analysis, the most energetic HV structures are located closer to the wall from the very beginning of the plate. Their wall-normal position continues to decrease downstream, albeit more gradually, as they are already close to the wall. Comparing AW and CW cases, LV structures exhibit similar amplitudes, but remain systematically located closer to the wall in the CW case (around $\eta\simeq 0.5$ compared to $\eta\simeq 0.6$ in AW). HV structures, on the other hand, display larger peak amplitudes in the AW configuration once the rapid growth phase is initiated, as observed for $p_c=0$.

The analysis of the instantaneous snapshots as those displayed in figure~\ref{fig:flow_description} has shown that breakdown systematically occurs within low-velocity streaks, as also reported in previous studies,  and is never  observed to initiate within high-velocity streaks. However, the sharp amplification and inward migration of high-velocity structures observed upstream of transition suggest that they also play an active role in the transition process.

    \subsubsection{Streak breakdown and turbulent spot formation \label{sec:streak_desta_and_turbulent_spot}}

\begin{figure}
    \centering

    \begin{subfigure}[t]{\linewidth}
        \centering
        \includegraphics[scale=1]{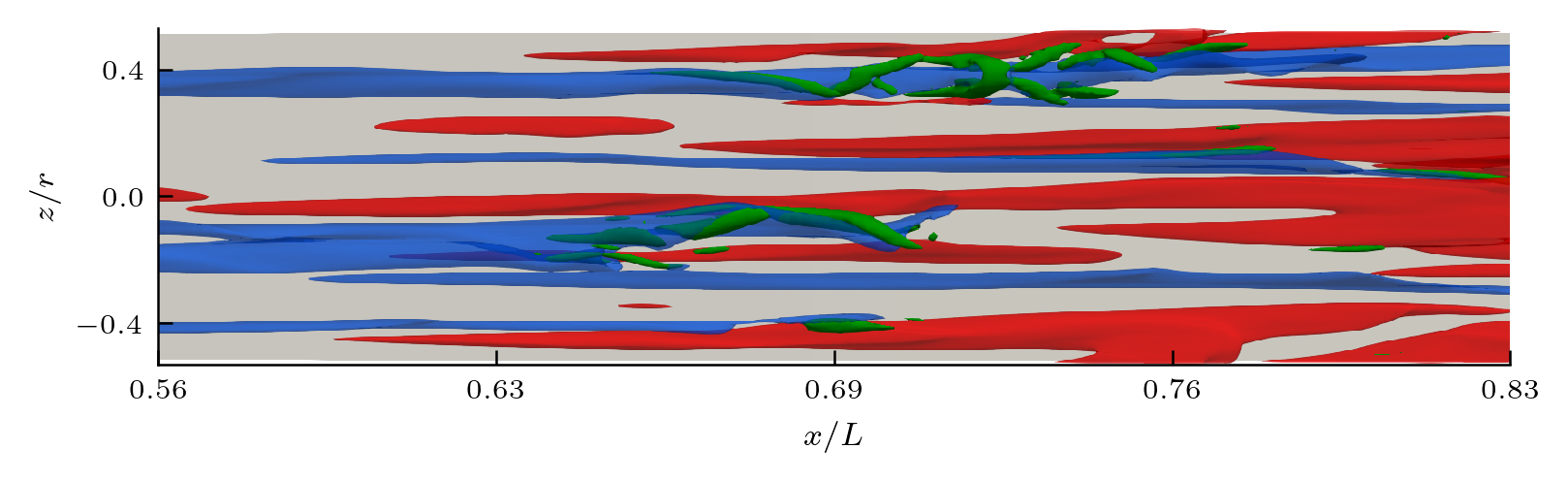}
        \caption{}
        \label{fig:instab_a}
    \end{subfigure}

    \vspace{0.3cm}

    \begin{subfigure}[t]{\linewidth}
        \centering
        \includegraphics[scale=1]{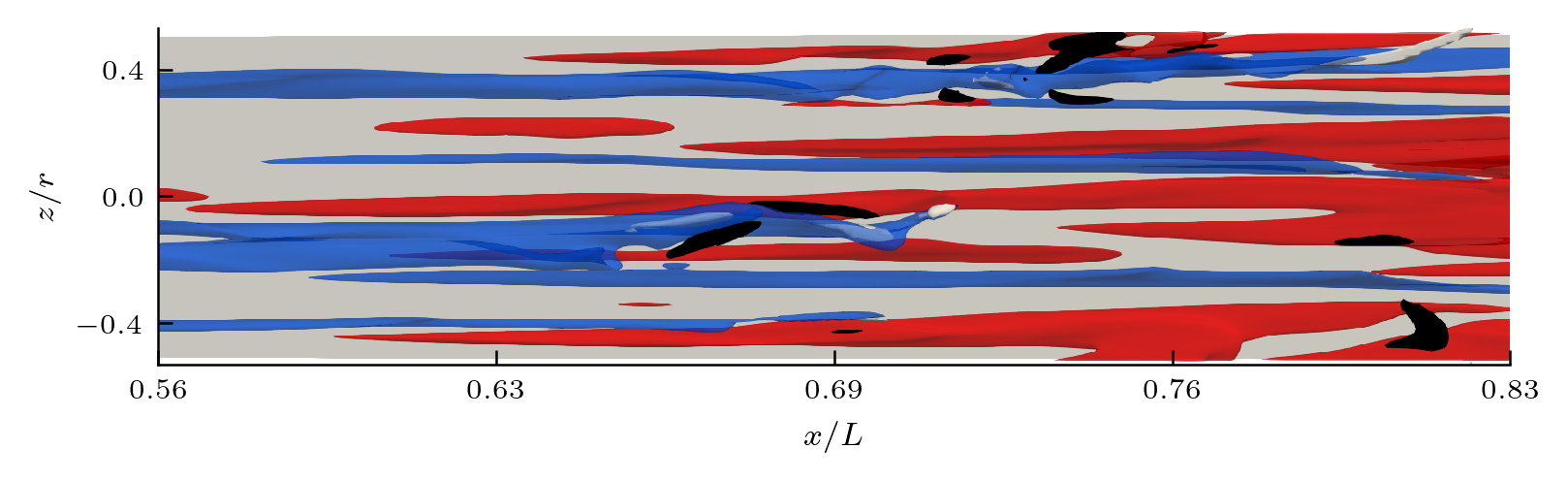}
        \caption{}
        \label{fig:instab_b}
    \end{subfigure}

    \vspace{0.3cm}

    \begin{subfigure}[t]{\linewidth}
        \centering
        \includegraphics[scale=1]{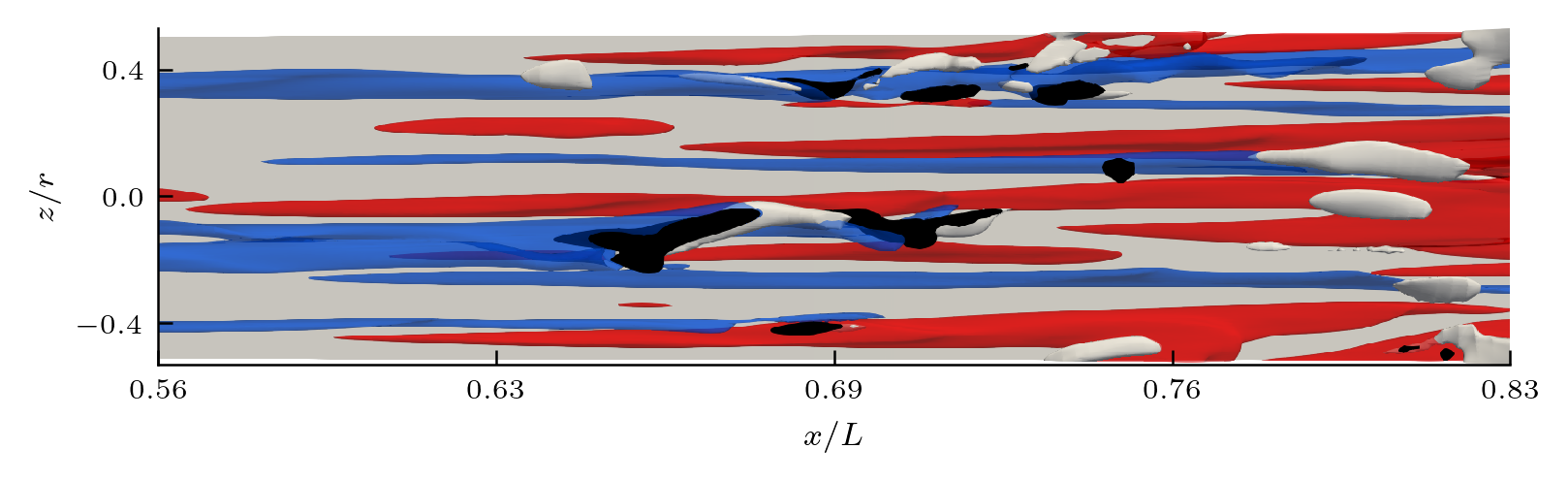}
        \caption{}
        \label{fig:instab_c}
    \end{subfigure}

    \caption{
    Snapshot from the CW simulation showing the simultaneous development of sinuous and varicose instabilities.
    In all panels, red and blue transparent isosurfaces correspond respectively to
    $u'_{\parallel} = 12\%\,U_{\infty}$ and $u'_{\parallel} = -12\%\,U_{\infty}$.
    (\textit{a}) Iso-surfaces of the adimensional $Q$-criterion, $Q^*=3$.
    (\textit{b}) Contours of $v'_{\perp}=-4.5\%\,U_{\infty}$ (black) and
    $v'_{\perp}=4.5\%\,U_{\infty}$ (white).
    (\textit{c}) Contours of $w'=-4.5\%\,U_{\infty}$ (black) and
    $w'=4.5\%\,U_{\infty}$ (white).
    }

    \label{fig:instability_modes}
\end{figure}

After an initial growth phase, a subset of streaks undergoes secondary instability. Both sinuous and varicose modes are observed in the AW and CW simulations. Figure~\ref{fig:instability_modes} presents a representative snapshot extracted from the CW case, in which a sinuous instability is visible at $z/r \simeq -0.1$ and a varicose instability at $z/r \simeq 0.4$. The two instability modes coexist within the same instantaneous flow field and develop on distinct low-velocity streaks.

\begin{figure}[htbp]
    \centering

    \begin{subfigure}{\textwidth}
        \centering
        \includegraphics[scale=1]{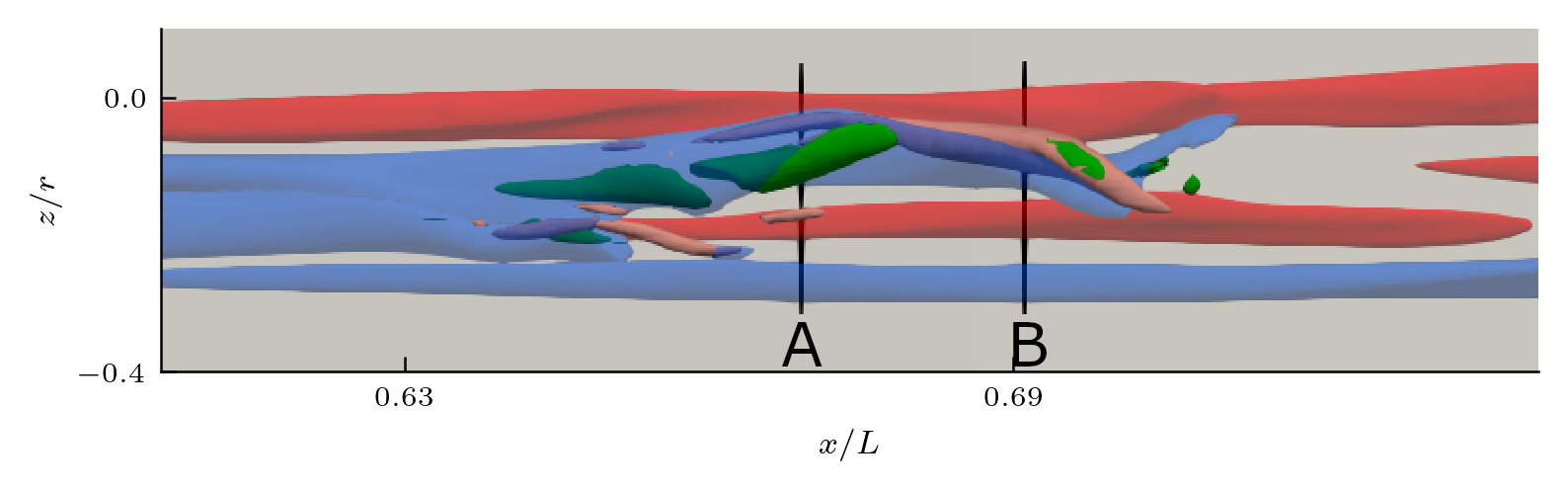}
        \caption{}
    \end{subfigure}

    \vspace{0.5cm}

    \begin{subfigure}{0.45\textwidth}
        \centering
        \includegraphics[scale=1]{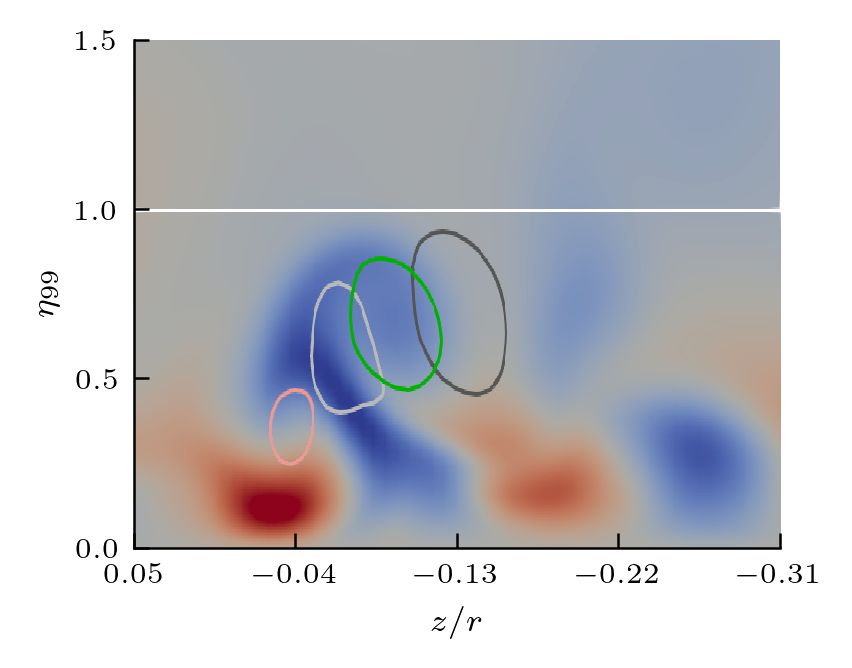}
        \caption{Section A;  $v'_{\perp}$ isocontours}
        \label{fig:sinuous:sectionA_v}
    \end{subfigure}
    \hfill
    \begin{subfigure}{0.45\textwidth}
        \centering
        \includegraphics[scale=1]{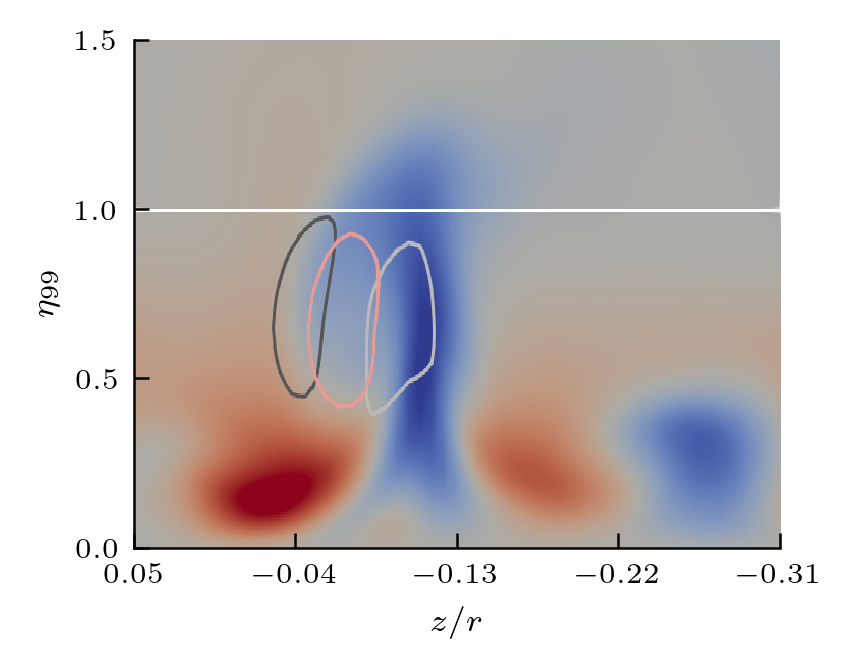}
        \caption{Section B;  $v'_{\perp}$ isocontours}
        \label{fig:sinuous:sectionB_v}
    \end{subfigure}

    \vspace{0.5cm}

    \begin{subfigure}{0.45\textwidth}
        \centering
        \includegraphics[scale=1]{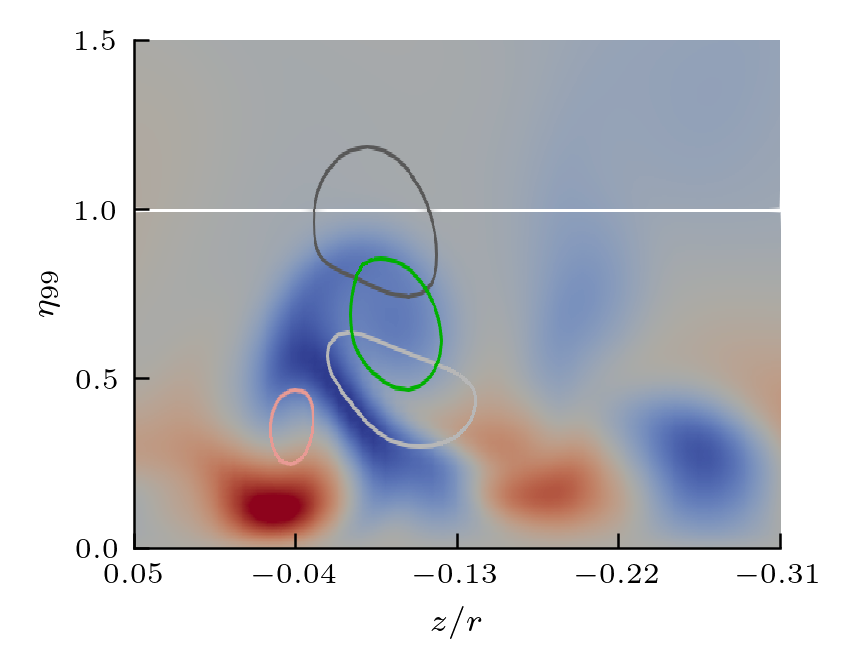}
        \caption{Section A;  $w'$ isocontours}
        \label{fig:sinuous:sectionA_w}
    \end{subfigure}
    \hfill
    \begin{subfigure}{0.45\textwidth}
        \centering
        \includegraphics[scale=1]{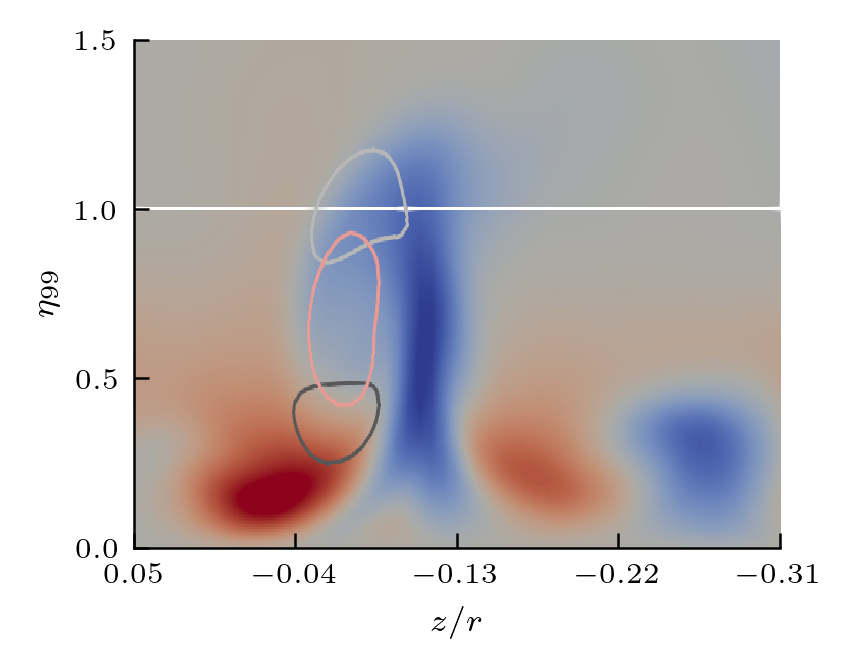}
        \caption{Section B;  $w'$ isocontours}
        \label{fig:sinuous:sectionB_w}
    \end{subfigure}

\caption{Visualisation of the isolated sinuous instability.
(\textit{a}) Three-dimensional view. Red and blue transparent isosurfaces correspond to
$u'_{\parallel}=12\%\,U_\infty$ and $u'_{\parallel}=-12\%\,U_\infty$, respectively.
Vortical structures identified by $Q^*=3$ are coloured by the sign of the streamwise vorticity,
with green denoting $\omega_x>0$ and pink denoting $\omega_x<0$.
The black lines indicate the locations of sections A and B.
(\textit{b,d}) Section A showing the $u'_{\parallel}$ field together with isocontours of $Q^*=3$ coloured by $\omega_x$. Isocontours of $v'_{\perp}=\pm 4.5\%$ are displayed in (\textit{b}) and isocontours of $w'=\pm 4.5\%$ in (\textit{d}) (white: positive, black: negative).
(\textit{c,e}) Same as (\textit{b,d}) for section B.
The horizontal white line indicates the local boundary-layer thickness $\delta_{99}$.
}
    \label{fig:sinuous}
\end{figure}
\begin{figure}[htbp]
    \centering

    \begin{subfigure}{\textwidth}
        \centering
        \includegraphics[scale=1]{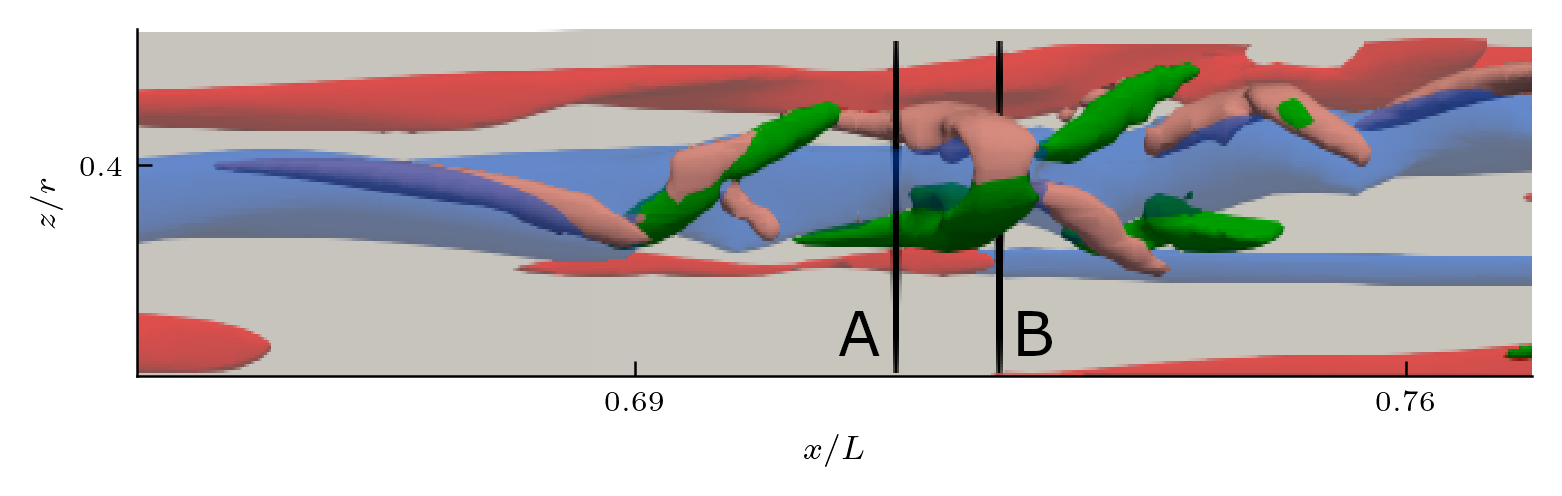}
        \caption{}
    \end{subfigure}

    \vspace{0.5cm}

    \begin{subfigure}{0.45\textwidth}
        \centering
        \includegraphics[scale=1]{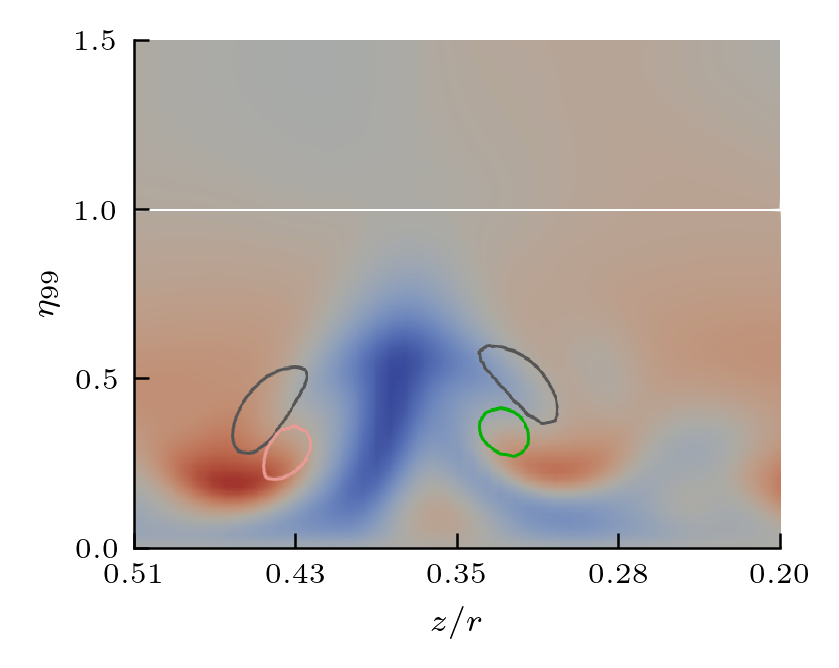}
        \caption{Section A;  $v'_{\perp}$ isocontours}
        \label{fig:varicose:sectionA_v}
    \end{subfigure}
    \hfill
    \begin{subfigure}{0.45\textwidth}
        \centering
        \includegraphics[scale=1]{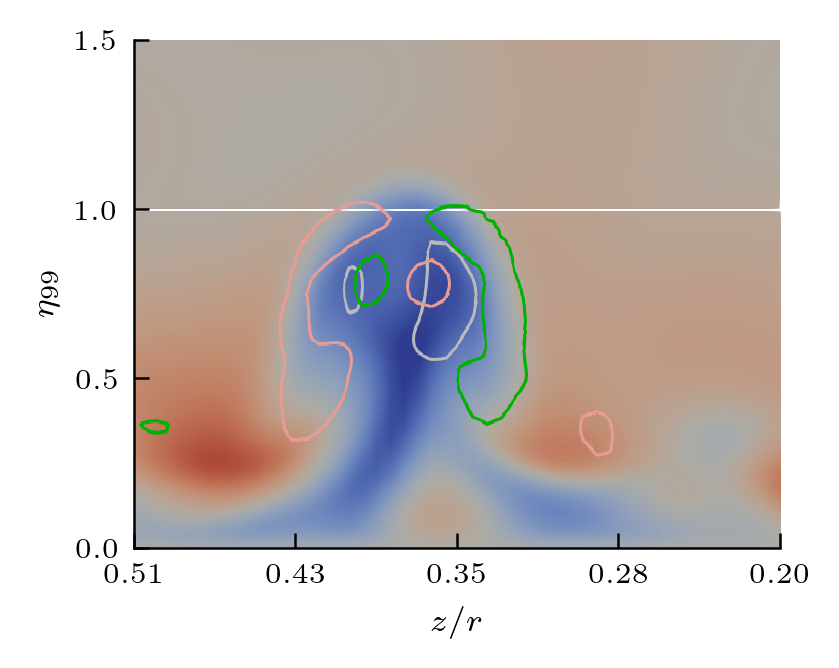}
        \caption{Section B;  $v'_{\perp}$ isocontours}
        \label{fig:varicose:sectionB_v}
    \end{subfigure}

    \vspace{0.5cm}

    \begin{subfigure}{0.45\textwidth}
        \centering
        \includegraphics[scale=1]{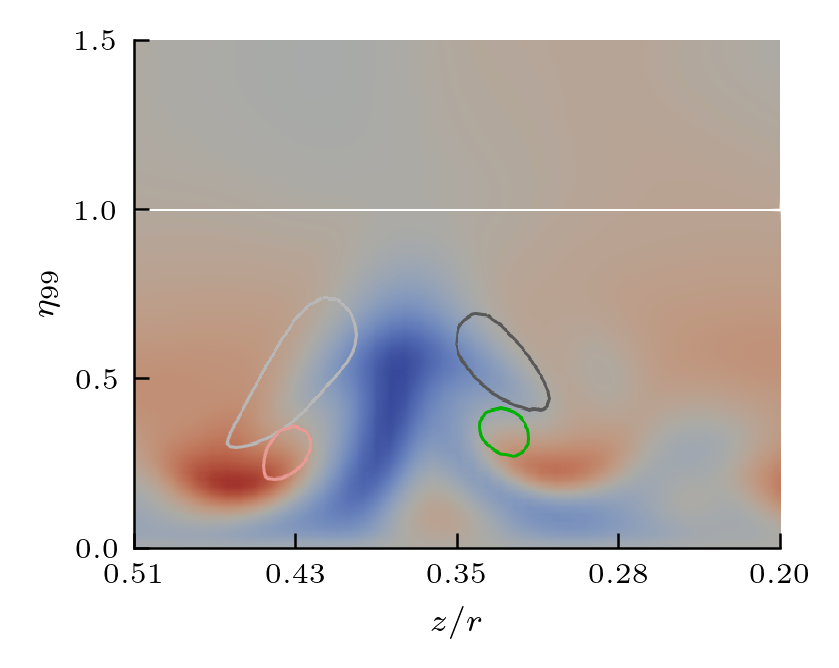}
        \caption{Section A;  $w'$ isocontours}
        \label{fig:varicose:sectionA_w}
    \end{subfigure}
    \hfill
    \begin{subfigure}{0.45\textwidth}
        \centering
        \includegraphics[scale=1]{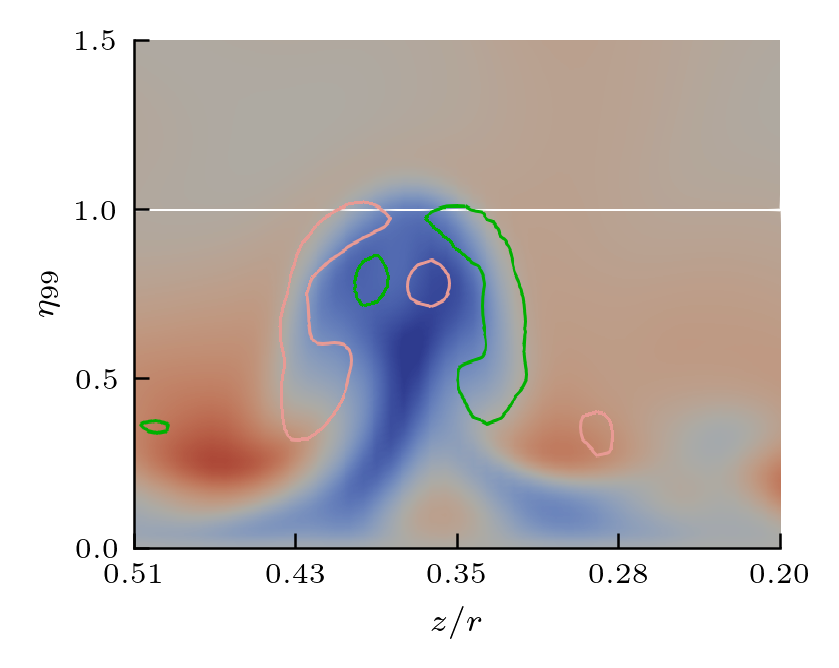}
        \caption{Section B;  $w'$ isocontours}
        \label{fig:varicose:sectionB_w}
    \end{subfigure}

    \caption{
    Visualisation of the isolated varicose instability. Same representation as in figure~\ref{fig:sinuous}. For section B, $Q^*$ is lowered to 0.2 to properly identify the hairpin vortex structure. 
    }
    \label{fig:varicose}
\end{figure}
The sinuous instability is characterised by the appearance of quasi-streamwise
vortices arranged in a staggered pattern on either side of the low-velocity
streak (\ref{fig:instab_a}). The associated wall-normal velocity fluctuation is
strongly asymmetric (\ref{fig:instab_b}), with upward motion concentrated
within the low-velocity streak and downward motion in the surrounding vortical
structures. In contrast, the spanwise velocity fluctuation exhibits a symmetric
organisation with respect to the streak (\ref{fig:instab_c}), with alternating
positive and negative fluctuations that induce a lateral oscillation of the
streak.
Figure~\ref{fig:sinuous} isolates the sinuous event and highlights
the organisation of the vortical structures through the sign of the streamwise
vorticity. The downstream (head) vortex carries negative streamwise vorticity
(pink) and develops between the unstable low-velocity streak and a strong
high-velocity streak located closer to the wall. Cross-section~B shows that
this vortex induces a positive wall-normal velocity at the streak centre
(\ref{fig:sinuous:sectionB_v}), lifting the low-velocity streak away from the
wall, while negative wall-normal velocity on its left side drives the adjacent
fluid downward. At the same time, the spanwise velocity field
(\ref{fig:sinuous:sectionB_w}) displaces the upper part of the streak towards
the left, whereas the near-wall portion is shifted towards the right, bringing
the neighbouring high-velocity streak closer to the low-velocity streak.
Further upstream, cross-section~A reveals a vortex of positive streamwise
vorticity (green) located on the opposite side of the streak. The wall-normal
motion remains qualitatively similar to that observed in section~B: positive
$v'_{\perp}$ lifts the streak core, while negative $v'_{\perp}$ pushes its
right-hand side towards the wall
(\ref{fig:sinuous:sectionA_v}). The spanwise displacement, however, is reversed:
the upper part of the streak is shifted towards the right, whereas the
near-wall region is displaced towards the left
(\ref{fig:sinuous:sectionA_w}).
Together, these staggered vortical structures generate the sinuous motion of
the low-velocity streak. As the streak is convected downstream alongside a
given vortex, its core is first lifted away from the wall and deflected
laterally. Further downstream, the same vortex progressively returns the streak
towards the wall while maintaining the lateral displacement. The next vortex,
offset in the streamwise direction and located on the opposite side of the
streak, repeats the same sequence with the opposite spanwise deflection. This
alternating succession of lifting, lateral displacement and return towards the
wall gives rise to the characteristic sinuous waviness of the streak.

The varicose instability exhibits a markedly different organisation. Vortical structures develop simultaneously on both sides of the low-velocity streak
(\ref{fig:instab_a}), resulting in a symmetric distribution of wall-normal
velocity fluctuation (\ref{fig:instab_b}) and an antisymmetric distribution of
spanwise velocity fluctuation (\ref{fig:instab_c}).
The isolated event shown in figure~\ref{fig:varicose} illustrates the mechanism
described by \citet{Brandt2004}. Counter-rotating vortices emerge
simultaneously on either side of the low-velocity streak, progressively
approach one another and eventually merge above the streak centreline. This
interaction generates inclined $\Lambda$- and V-shaped vortical structures,
which subsequently evolve into hairpin vortices composed of two
counter-rotating legs.
Cross-section~A intersects the two vortices before their
merging. The left vortex carries negative streamwise vorticity (pink), whereas
the right vortex carries positive streamwise vorticity (green), both being
located at the interface between the low-velocity streak and the neighbouring
high-velocity streaks. Each vortex induces downward wall-normal motion above
its core (\ref{fig:varicose:sectionA_v}), producing a symmetric downward
displacement on both sides of the low-velocity streak. Since the vortices are
located in the lower part of the boundary layer
($\eta_{99}\approx0.3$), this motion maintains the streak confined within the
boundary layer. At the same time, the opposite signs of streamwise vorticity
generate opposite spanwise motions that spread the streak laterally
(\ref{fig:varicose:sectionA_w}). The combined action of downward displacement
and lateral spreading therefore broadens the low-velocity streak while preventing it from spreading outside of the boundary layer, as occurs for the sinuous instability. 
Further downstream, cross-section~B intersects the developing hairpin vortex (\ref{fig:varicose:sectionB_v}, \ref{fig:varicose:sectionB_w}).
The two larger vortices associated with the hairpin structure remain located on
either side of the low-velocity streak, joining each other at its top at the boundary layer edge, while the two smaller vortices near the
streak centre correspond to the upstream V-shaped structure. At this streamwise
location, the section lies between the extrema of the wall-normal and spanwise
velocity fluctuations visible in
figures~\ref{fig:instab_b} and~\ref{fig:instab_c}. Consequently, the vortical
organisation is clearly identifiable, whereas the associated wall-normal and
spanwise velocity signatures are locally weak.

Overall, both sinuous and varicose secondary instabilities are observed in the CW simulation, showing that strong wall cooling does not fundamentally alter these mechanisms. Beyond the differences between the two instability modes, a common cross-sectional organisation is observed once the streak deformation becomes pronounced. The upper part of the low-velocity streak undergoes significant lateral spreading, whereas its lower part remains confined between the adjacent high-velocity streaks, giving the streak an overall mushroom-like shape. This behaviour is consistent with the preferential amplification of the neighbouring high-velocity streaks in the near-wall region, which progressively occupy the available space close to the wall and thereby constrain the deformation of the low-velocity streak. 

In both LES (AW and CW), secondary instability consistently develops on low–velocity streaks, subsequently leading to spot formation. Some events occur close to the leading edge, so spots appear early in the domain.
To study them, the downstream convection of turbulent spots was characterised by tracking the 
positions of their leading and trailing edges over time on both the top and bottom 
sides of the plate. For each spot, the instantaneous edge positions were identified, 
and a linear regression was performed to extract a constant convection velocity. 
        \begin{figure}
            \centering
            \begin{subfigure}[t]{\linewidth}
                \centering
                \includegraphics[scale=1]{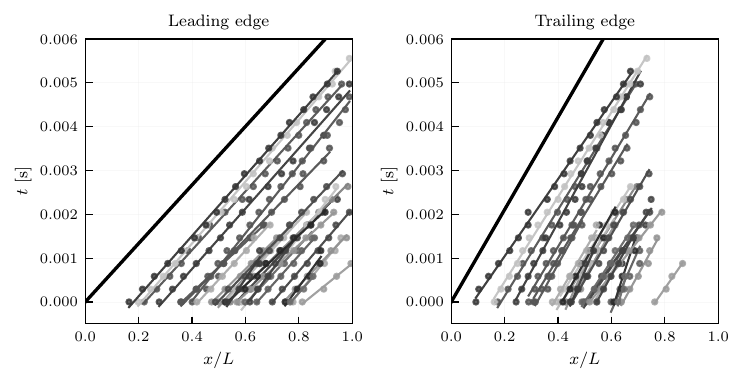}
                \caption{AW}
                \label{fig:LE_TE_adia}
            \end{subfigure}%
            \hfill
            \begin{subfigure}[t]{\linewidth}
                \centering
                \includegraphics[scale=1]{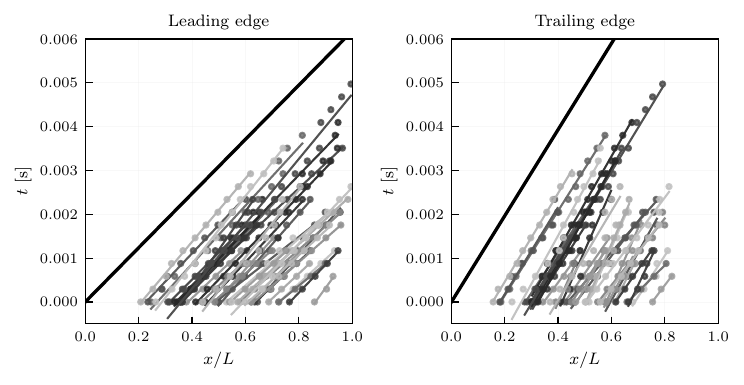}
                \caption{CW}
                \label{fig:LE_TE_PF}
            \end{subfigure}
            \caption{
              Convection of turbulent spots: temporal evolution of the leading-edge and trailing-edge positions for the AW (\textit{a}) and CW (\textit{b}) configurations.
              Symbols indicate individual measurements on the top and bottom sides of the plate, 
              while grey lines correspond to linear fits used to estimate the edge convection velocities from those measurements. 
              The thick black line in each panel shows the ensemble-averaged fit. 
              }
            \label{fig:spots_convection}
        \end{figure}
The scatter plots in figure~\ref{fig:spots_convection} illustrate the measured 
trajectories together with their corresponding linear fits, from which the edge 
velocities were obtained. 
In the AW case, the leading edge propagates downstream at a speed of 
approximately $0.87\,U_\infty$, while the trailing edge follows more slowly at 
about $0.55\,U_\infty$, both values corresponding to ensemble averages over all 
detected spots. In the CW case, higher velocities are observed, 
$0.94\,U_\infty$ for the leading edge and $0.59\,U_\infty$ for the trailing edge. The simulations thus recover the characteristic leading- and trailing-edge convection velocities measured experimentally by \citet{Mans2005}.
This systematic difference between the two edges reflects the continuous stretching 
of turbulent spots as they are advected downstream by the outer flow. Furthermore, 
figure~\ref{fig:spots_convection} shows that in both the AW and CW cases, 
turbulent spots emerge as early as $x/L=0.2$. More spots were detected in the CW 
configuration (36) than in the AW case (27). The birth of new spots is distributed 
fairly uniformly over the range $0.2 \leq x/L \leq 1$, with no preferential 
streamwise location for their onset.

\section{Discussion and conclusions}

We have performed and analysed two wall-resolved, time-resolved LES of bypass transition over a flat plate with an elliptical leading edge, under adiabatic-wall (AW) and cooled-wall (CW) conditions. Retaining the leading edge allows the boundary layer to develop naturally from the stagnation region under incoming FST, so that receptivity, streak formation and subsequent breakdown arise without artificial forcing inside the boundary layer. The geometry is sized for experimental reproducibility, including the cooled-wall configuration, thereby providing a framework for future numerical--experimental comparisons.
The simulations capture the full receptivity pathway in both cases, resolving the vortex-tilting and lift-up mechanisms responsible for the generation of streamwise streaks. The wall-normal transport term \(v'_{\perp}\partial_n U_{\parallel}\) is identified as a clear precursor of streak formation, while the shear-sheltering effect is quantified by directly linking the penetration depth of external disturbances to their frequency content.

Wall cooling does not alter the nature of the transition, which follows the canonical bypass route in both configurations, but it does modulate several flow features. In CW, thermal streaks accompany the streamwise-velocity streaks; they are generated by the same lift-up process and remain tightly correlated with their velocity counterparts. The velocity streaks reside lower in the boundary layer than in AW while exhibiting similar amplitudes, streamwise growth rates and spanwise spacings, thereby confirming the wallward displacement predicted by optimal perturbation theory. The CW simulation also recovers the two classical secondary-instability modes reported in the bypass-transition literature, namely sinuous and varicose streak instabilities. Taken together, these results demonstrate that wall cooling modifies the thermal and spatial organisation of the streaks while preserving the fundamental mechanisms governing their amplification and breakdown. The onset location of transition remains essentially unchanged, yet the subsequent progression towards turbulence is slightly accelerated under wall cooling. Consistently, turbulent spots convect faster in CW than in AW, which may contribute to the more rapid completion of transition.

The conditional analysis establishes a clear asymmetry between high- and low-velocity streaks throughout the bypass-transition process. 
In the pre-transitional regime, low-velocity streaks are located farther from the wall and generally exhibit larger amplitudes than high-velocity streaks.
The present results further show that the strongest evolution preceding transition is not associated with the low-velocity streaks themselves, but rather with the high-velocity streak population. Indeed, the rapid amplification and displacement of high-velocity structures towards the wall begin upstream of transition onset. Examination of the instantaneous flow fields nevertheless shows that breakdown is systematically initiated within low-velocity streaks and is never observed within high-velocity streaks. High-velocity streaks therefore do not appear to constitute the direct site of instability onset.
Instead, the observed behaviour suggests that high-velocity streaks participate in a broader reorganisation of the streak field preceding breakdown, potentially through the nonlinear lift-up mechanism proposed by \citet{Mao2017}, in which nonlinear streak interactions progressively lift low-velocity streaks towards the outer part of the boundary layer while displacing high-velocity streaks towards the wall. The amplification and progressive confinement towards the near-wall region of high-velocity streaks increase the velocity contrast between neighbouring high- and low-velocity streaks and may strengthen the local shear layers separating them. Such a scenario supports the interpretation proposed by \citet{Nolan2012}, who argued that bypass transition is primarily governed by interactions between streaks within the boundary layer rather than by direct penetration of free-stream disturbances from the outer flow. The evolution of the high-velocity streak population may thus contribute indirectly to the conditions leading to secondary instability, even though the instability itself develops within low-velocity streaks.
It is also worth noting that the conditional averages presented here characterise the behaviour of the overall streak population. While the average low-velocity streak remains located within the boundary layer throughout the pre-transitional regime, the instantaneous streaks that undergo breakdown may experience substantially larger wall-normal displacement than suggested by the conditional statistics.

Although the present plate length does not allow the complete establishment of a fully turbulent boundary layer, it captures the entire receptivity-to-breakdown sequence that characterises bypass transition. Within this framework, the present results identify the evolution of the high-velocity streak population as an important aspect of the pre-transitional dynamics that deserves further investigation. Finally, the experimentally reproducible nature of the present configuration opens the possibility of future detailed comparisons between measurements and simulations, offering a valuable framework for further investigation of wall-temperature effects and streak dynamics during bypass transition.

\section*{Acknowledgements} 
This project was provided with computer and storage resources by GENCI at CINES thanks to grants A0052A10589 and A0082A10589 on the supercomputer Occigen. The authors also gratefully acknowledge L. Jecker for preliminary post-processing work on the simulations.

During the
preparation of this manuscript, the corresponding author used ChatGPT (GPT-5.6 Sol, OpenAI) to assist with English-language editing
and the reformulation of author-written text. All AI-assisted text was
reviewed and revised by the authors, who take full responsibility for
the final content.

\section*{Declaration of interests} 
The authors report no conflict of interest.

\section*{Author contributions}
A. Veilleux: Formal analysis, Visualization, Investigation, Data curation, Writing – original draft, Writing – review \& editing.
H. Deniau: Methodology, Software, Investigation, Formal analysis, Writing – review \& editing. 
O. Vermeersch: Conceptualization, Methodology, Data curation, Resources.

\begin{appen}

\section{Reference homogeneous turbulence simulation}
 \label{app:cube}

To isolate the behaviour of the synthetic-turbulence forcing from the
geometrical effects of the flat-plate configuration, an auxiliary LES was
performed in a homogeneous rectangular box using the same solver, forcing parameters and
target energy spectrum as in the present study.
The computational domain extends over
$0.10\times0.05\times0.05$~m$^3$ and is discretised with an isotropic Cartesian
mesh ($\Delta x=\Delta y=\Delta z\simeq 10^{-4}$~m). Periodic boundary
conditions are applied in the transverse directions, while turbulence is
injected at the inlet and convected through the domain without any solid
boundary.
Energy spectra evaluated from the velocity signal recorded at the location $x=0.03$~m, $y=0.025$~m, $z=0.025$~m are presented in figure~\ref{fig:spectre_cube}.
The three velocity-component spectra remain of comparable magnitude throughout
the energetic part of the resolved frequency range, and no component exhibits a
systematic dominance. This indicates that the synthetic-turbulence forcing does
not introduce a significant intrinsic directional bias. 
The total spectrum reproduces the general shape and characteristic
energy-containing range predicted by the theoretical spectrum obtained from
Pope's model, although a progressively increasing energy
deficit is observed at high frequencies, particularly for
$f \gtrsim 10^4$~Hz. This faster spectral decay is attributed to the combined
effects of numerical filtering and small-scale dissipation. The auxiliary case
is therefore used here primarily to assess the directional balance of the
forcing, rather than to establish an exact spectral agreement over the entire
resolved frequency range.

      \begin{figure}
          \centering 
        \includegraphics[scale=1]{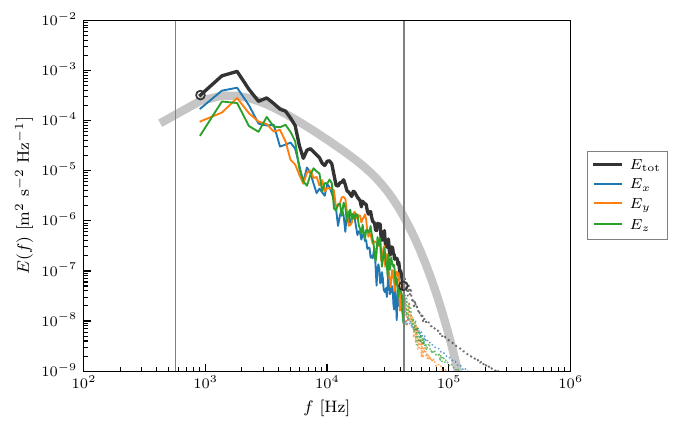}
        \caption{
        Energy spectra of the velocity fluctuations in the auxiliary homogeneous-box
        simulation. Other plotting conventions are identical to those of
        figure~\ref{fig:spectrum_Pope_PF}. The spectra are evaluated at
        $(x,y,z)=(0.03,\,0.025,\,0.25)$~m.
        }
        \label{fig:spectre_cube}
      \end{figure}

\section{Post-processing methodology}
\label{app:post}

Three categories of outputs were generated during the simulations and used in the present analysis:
(i) time-averaged databases accumulated during the LES,
(ii) stored instantaneous fields, and
(iii) probe signals.
For consistency with the plate geometry (elliptical leading edge), quantities are expressed in a local curvilinear frame $(t,n,z)$, where $t$ and $n$ denote the unit tangent and unit normal to the surface, and $z$ is the spanwise direction. Tangential and normal components are denoted by the subscripts $\parallel$ and $\perp$, respectively.

\subsection{Time-averaged turbulence statistics}
\label{app:stats}

First- and second-order moments in the Cartesian frame are accumulated during the LES, from which centered moments are reconstructed. The Reynolds stresses are then projected onto the local wall-based frame. Denoting by $\mathbf{t}=(t_x,t_y,0)$ the unit tangent vector and $\mathbf{n}=(-s\,t_y,s\,t_x,0)$ the outward unit normal, where \(s=\pm1\)
accounts for the wall orientation, the streamwise-aligned and wall-normal fluctuations are defined as
\begin{equation}
u_\parallel' = t_x u' + t_y v',\qquad
v_\perp' = s(-t_y u' + t_x v').
\end{equation}

The corresponding second-order moments in the wall-based frame are obtained by rotation of the Reynolds-stress tensor,
\begin{equation}
\langle u_\parallel'^2\rangle
= t_x^2\langle u'^2\rangle + 2 t_x t_y \langle u'v'\rangle + t_y^2 \langle v'^2\rangle,
\end{equation}
\begin{equation}
\langle v_\perp'^2\rangle
= t_y^2\langle u'^2\rangle - 2 t_x t_y \langle u'v'\rangle + t_x^2 \langle v'^2\rangle,
\end{equation}
with similar expressions for the cross-correlations. The root-mean-square (rms) levels are then defined as
\begin{equation}
u_{\parallel,\mathrm{rms}} = \sqrt{\langle u_\parallel'^2\rangle},\qquad
v_{\perp,\mathrm{rms}} = \sqrt{\langle v_\perp'^2\rangle}.
\end{equation}

\subsection{Instantaneous fields}
\label{app:inst}

Instantaneous velocity fields $\{u(\mathbf{x},t^k),v(\mathbf{x},t^k),w(\mathbf{x},t^k)\}$ are stored at discrete times $t^k$. Fluctuations are obtained by subtracting the global mean fields accumulated during the LES,
\begin{equation}
u'(\mathbf{x},t^k)=u(\mathbf{x},t^k)-\langle u\rangle(\mathbf{x}), \quad \text{etc.}
\end{equation}
and are projected onto the local wall-based frame as
\begin{equation}
u_\parallel'(\mathbf{x},t^k)=t_x u'(\mathbf{x},t^k)+t_y v'(\mathbf{x},t^k),
\end{equation}
\begin{equation}
v_\perp'(\mathbf{x},t^k)=s\!\left[-t_y u'(\mathbf{x},t^k)+t_x v'(\mathbf{x},t^k)\right].
\end{equation}

Second-order statistics are then obtained by averaging over the $N$ stored snapshots,
\begin{equation}
\langle u_\parallel'^2\rangle(\mathbf{x})
= \frac{1}{N} \sum_{k=1}^{N} \left[u_\parallel'(\mathbf{x},t^k)\right]^2,
\end{equation}
with analogous expressions for all components and correlations.

By construction, this procedure yields the same quantity $u_{\parallel,\mathrm{rms}}  = \sqrt{\langle u_\parallel'^2\rangle}$ as that obtained from the time-averaged LES statistics, but based on a finite set of snapshots.

\subsection{Probe signals}
\label{app:probes}
At each probe location, time series of the velocity components are recorded. The same post-processing procedure as described for the instantaneous fields is applied, the only difference being the temporal sampling and the absence of spatial averaging.

Fluctuations are defined with respect to the local temporal mean at each probe,
\begin{equation}
u'(t)=u(t)-\langle u\rangle_t,\qquad
v'(t)=v(t)-\langle v\rangle_t,\qquad
w'(t)=w(t)-\langle w\rangle_t,
\end{equation}
where $\langle \cdot \rangle_t$ denotes time averaging over the recorded signal.

The fluctuation components are projected onto the local wall-based frame $(t,n,z)$ as
\begin{equation}
u_\parallel'(t)=t_x u'(t)+t_y v'(t),\qquad
v_\perp'(t)=s\!\left[-t_y u'(t)+t_x v'(t)\right],
\end{equation}
while the spanwise component remains unchanged.

Second-order moments are obtained from temporal averaging,
\begin{equation}
\langle u_\parallel'^2\rangle_t = \langle u_\parallel'(t)^2\rangle_t,
\end{equation}
with analogous definitions for all components and correlations. The corresponding rms levels are then defined as
\begin{equation}
u_{\parallel,\mathrm{rms}} = \sqrt{\langle u_\parallel'^2\rangle_t},\qquad
v_{\perp,\mathrm{rms}} = \sqrt{\langle v_\perp'^2\rangle_t}.
\end{equation}

 \section{Intermittency detection method}
 \label{app:intermittency}
The intermittency factor was obtained from the LES database using the detector function based on cross-stream velocity fluctuations proposed by \citet{Nolan2013}. Instantaneous flow fields were stored every 7520 iterations, and $(x,z)$ slices were extracted at several fixed wall-normal positions corresponding to $\eta=0.1$--$0.8$, where $\eta=y/\delta_{99}$. At each slice, the local detector was defined as the sum of the magnitudes of the velocity fluctuations in the $y$- and $z$-components, normalised by their spatial standard deviation. A binary value of~1 (turbulent) was assigned when the local fluctuation level exceeded the global standard deviation of the slice, and 0 otherwise. The resulting binary field was then averaged in the spanwise direction and over all stored snapshots, yielding the intermittency factor $\gamma(x)$ at each wall-normal position. 

The raw intermittency values obtained from the detector  are shown as red symbols in figure~\ref{fig:method_intermittency}. Close to the leading edge, non-zero intermittency values are observed ($\gamma \approx 0.1$ at $x/L=0$), which are attributed to turbulent fluctuations from the free-stream intermittently penetrating the boundary layer. These values are not representative of the intrinsic transition process and are therefore not considered in the fit. In practice, the raw intermittency first reaches a small maximum near the leading edge, then decreases and attains a minimum further downstream. Beyond this point, the influence of external turbulence vanishes and the subsequent rise of $\gamma$ genuinely reflects the beginning of transition. The sigmoidal approximation is thus constructed using only the points located downstream of this minimum. For completeness, the upstream points are retained in the plot but their values are replaced by the minimum level, ensuring that the fitted curve satisfies $\gamma$ close to zero at the inflow. The resulting data set is shown in black in figure~\ref{fig:method_intermittency}, together with the fitted sigmoidal law in blue.

    \begin{figure}
      \centerline{\includegraphics[scale=1]{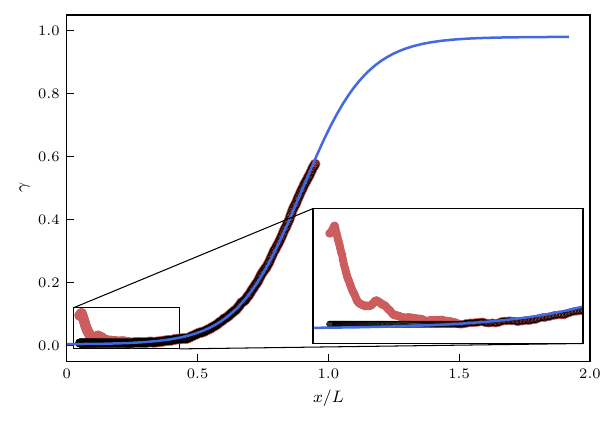}}
      \caption{Illustration of the data selection procedure for the sigmoidal fit of the intermittency. Raw values from the detector are shown in red. Only the points located downstream of the minimum are used to construct the fit, while the upstream values are set to the minimum level (black). The resulting sigmoidal approximation is shown in blue.}
      \label{fig:method_intermittency}
    \end{figure}

 \section{Validation of the snapshot-based statistics\label{app:validation_MC_snapshots}}

Figure~\ref{fig:valid_MC} assesses the convergence of the snapshot-based statistics by comparing them to reference quantities obtained from time-averaged fields over the entire simulation. The streamwise evolution of the streak amplitude,
\begin{equation}
A(x) = \max_{\eta<1} \left( u_{\parallel,\mathrm{rms}} \right) / U_\infty,
\end{equation}
and of its wall-normal position $\eta_A$, defined as the location of this maximum within the boundary layer, are reported for both the AW and CW cases. Results are shown for the upper (left) and lower (right) sides of the plate. For each configuration, the solid lines correspond to statistics computed from the fully converged mean fields, while dashed lines denote quantities extracted from the snapshot database. Despite the reduced temporal sampling (every 7520 iterations), a really close agreement is observed for both $A(x)$ and $\eta_A$, over the entire plate. This confirms that the snapshot frequency is sufficient to study the evolution of the streaks and validates the use of the conditional averaging procedure in the main analysis.
      \begin{figure}
          \centering 
      \begin{subfigure}{0.49\textwidth}
        \includegraphics[scale=1]{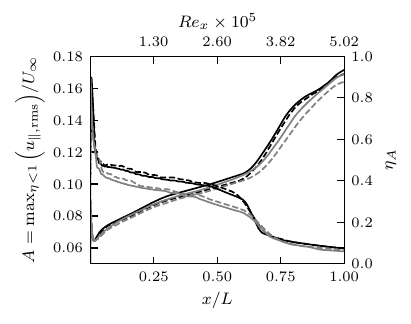}
        \caption{}
        \label{fig:valid_MC_bot}
      \end{subfigure}\hfil 
      \begin{subfigure}{0.49\textwidth}
        \includegraphics[scale=1]{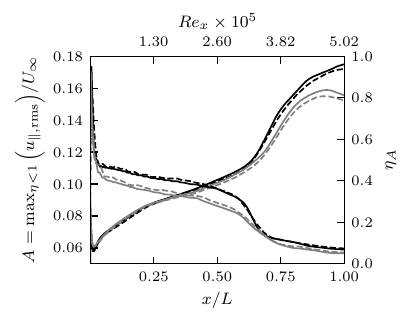}
        \caption{}
        \label{fig:valid_MC_top}
      \end{subfigure}

\vspace{0.5em}

\begin{center}
{\footnotesize
\setlength{\fboxsep}{4pt}
\setlength{\fboxrule}{0.3pt}

\fcolorbox{black!50}{white}{%
\begin{tabular}{
    c
    @{\hspace{1.2em}} c
    @{\hspace{1.5em}} c
}

& Turb. stat. & Snapshots \\[0.15em]

AW
&
\raisebox{0.3ex}{%
\tikz{
    \draw[
        line width=1.2pt
    ] (0,0) -- (0.70,0);
}}
&
\raisebox{0.3ex}{%
\tikz{
    \draw[
        line width=1.2pt,
        dash pattern=on 4pt off 2pt
    ] (0,0) -- (0.70,0);
}}
\\[0.2em]

CW
&
\raisebox{0.3ex}{%
\tikz{
    \draw[
        gray,
        line width=1.2pt
    ] (0,0) -- (0.70,0);
}}
&
\raisebox{0.3ex}{%
\tikz{
    \draw[
        gray,
        line width=1.2pt,
        dash pattern=on 4pt off 2pt
    ] (0,0) -- (0.70,0);
}}

\end{tabular}%
}}
\end{center}

\vspace{0.2em}

      \caption{Downstream evolution of the streak amplitude (left axis) and its wall-normal position (right axis), obtained from time-averaged
turbulence statistics and the snapshot database for the AW and CW
configurations. Results are shown separately for the suction side (a) and pressure side (b) of the plate. }

      \label{fig:valid_MC}
      \end{figure}
\end{appen}\clearpage

\bibliographystyle{jfm}
\bibliography{jfm}

\end{document}